\documentclass[final,5p,times,twocolumn,authoryear]{elsarticle}

\usepackage{amstext,amssymb,aas_macros,graphicx,epsfig}
\usepackage[breaklinks,colorlinks,pdfa=true]{hyperref}

\usepackage{tikz}
\usetikzlibrary{shapes.geometric, arrows}

\journal{Astronomy $\&$ Computing}

\begin{document}

\begin{frontmatter}
   \title{A GPU-accelerated viewer for HEALPix maps\smallskip\\
          \normalsize\url{https://github.com/andrei-v-frolov/healpix-viewer/}\\
          \normalsize\url{https://www.sfu.ca/physics/cosmology/healpix/}}
   \author[first]{Andrei V.\ Frolov}
   \ead{frolov@sfu.ca}
   \affiliation[first]{organization={Department of Physics, Simon Fraser University},
                addressline={8888 University Drive}, 
                city={Burnaby},
                postcode={V5A 1S6}, 
                state={BC},
                country={Canada}}

  \begin{abstract}
   {HEALPix by \citet{2005ApJ...622..759G} is a de-facto standard for
   Cosmic Microwave Background (CMB) data storage and analysis, and is widely
   used in current and upcoming CMB experiments. Almost all the datasets in
   Legacy Archive for Microwave Background Data Analysis
   (\href{https://lambda.gsfc.nasa.gov}{LAMBDA}) use HEALPix as a format
   of choice. Visualizing the data plays important role in research, and
   several toolsets were developed to do that for HEALPix maps, most notably
   original Fortran facilities and Python integration with \texttt{healpy}.}
   {With the current state of GPU performance, it is now possible to visualize
   extremely large maps in real time on a laptop or a tablet. HEALPix Viewer
   described here is developed for macOS, and takes full advantage of GPU
   acceleration to handle extremely large datasets in real time. It compiles
   natively on Intel and Arm64 architectures, and uses Metal framework for
   high-performance GPU computations. The aim of this project is to reduce
   the effort required for interactive data exploration, as well as time
   overhead for producing publication-quality maps. Drag and drop integration
   with Keynote and Powerpoint makes creating presentations easy.}
   {The main codebase is written in Swift, a modern and efficient compiled
   language, with high-performance computing parts delegated entirely to GPU,
   and a few inserts in C interfacing to \texttt{cfitsio} library for I/O.
   Graphical user interface is written in SwiftUI, a new declarative UI
   framework based on Swift. Most common spherical projections and colormaps
   are supported out of the box, and the available source code makes it easy
   to customize the application and to add new features if desired.}
   {On a M1 Max laptop, an $\texttt{nside}=8192$ maps are processed in real time,
   with geometry effects rendered at 60fps in full resolution with no
   appreciable load to the machine. Main user-facing delays are limited to
   CPU-bound \texttt{cfitsio} load times, and sorting needed to construct
   Cumulative Distribution Function (CDF) estimators for statistical analysis
   (hidden in background queue). Overall performance improves on the current
   Python software stack by a factor of 3-180x depending on the task at hand.}
  \end{abstract}
  
  

\end{frontmatter}

\newcommand{\atan}{\textrm{atan2}}
\newcommand{\code}[1]{\texttt{#1}}
\renewcommand{\vec}[1]{\ensuremath\mathbf{#1}}

\section{Introduction}
\label{sec:intro}

Since its original discovery \citep{1965ApJ...142..414D}, cosmic microwave background (CMB) has been a veritable gold mine of information on cosmology in general and early universe physics in particular. After COBE measured CMB spectrum to be almost perfectly that of black body radiation \citep{1990ApJ...354L..37M} and detected CMB temperature anisotropy for the first time \citep{1992ApJ...396L...1S}, implications for cosmology were quickly explored, e.g.\ see \cite{1992MNRAS.258P...1E}. Polarization of CMB was detected 10 years later \citep{2002Natur.420..772K}, and the field was maturing quickly with a number of ground, balloon-borne, and satellite experiments, as well as work by theorists, to become truly a source for precision cosmology. Two of the satellite missions, NASA's WMAP \citep{2013ApJS..208...20B} and ESA's Planck \citep{2020A&A...641A...1P} were particularly impactful. Current, future, and proposed CMB experiments will carry the torch, such as Simons Obervatory \citep{2019JCAP...02..056A}, LiteBIRD \citep{2012SPIE.8442E..19H, 2014JLTP..176..733M}, CMB Stage 4 \citep{2019arXiv190704473A}, and PICO \citep{2019arXiv190210541H}. Cosmic microwave background observations were used not only to measure cosmological parameters \citep{2013ApJS..208...19H, 2020A&A...641A...6P} and explore constraints on primordial fluctuations \citep{2020A&A...641A...7P, 2020A&A...641A...9P, 2020A&A...641A..10P}, but they also provide information on astrophysical foregrounds and are becoming important for galactic science, e.g.\ as discussed by \cite{2022ApJ...929..166H}.

De-facto standard for CMB data storage and analysis is HEALPix (Hierarchical Equal Area isoLatitude Pixelation) by \cite{1999astro.ph..5275G, 2005ApJ...622..759G, 2011ascl.soft07018G}, used by both WMAP and Planck \citep{2015IAUGA..2247779H}, as well as by most of the data published on LAMBDA (Legacy Archive for Microwave Background Data Analysis) \citep{2019arXiv190508667A}. Original HEALPix library and facilities were written in Fortran (with C bindings available), but it since has been ported or interfaced to a number of languages including R \citep{2019arXiv190705648F}, Julia \citep{2021ascl.soft09028T}, and most notably Python with \href{https://healpy.readthedocs.io/}{\texttt{healpy}} package \citep{2020ascl.soft08022Z}.

Scientific data visualization plays an important role in research, and two of the most widely used tools for visualizing HEALPix maps are original \texttt{map2gif} facility of HEALPix, and \texttt{healpy} package. A number of interactive 3D visualization tools were developed, either written in Java \citep{2008ASPC..394..319J, 2010ASPC..434..163F}, cross-platform frameworks such as as LAMBDA's \href{https://github.com/nasa-lambda/skyviewer/}{SkyViewer} (uses Qt) and Univiewer \citep{2010AstBu..65..296M} (uses wxWidgets), or targeting specific operating systems such as CMBview for macOS \citep{2011ascl.soft12011P}. Some of these projects appear to be orphaned, and GPU technology has come a long way since then, both in hardware and software available. OpenCL superseded OpenGL for massively parallel heterogeneous computing, and in turn was followed by more optimized but vendor-specific GPU libraries such as Apple's Metal or AMD's Vulkan. It is now possible to process extremely large datasets on commodity hardware, and even on portable devices.

This paper describes HEALPix Viewer, a macOS application I wrote to visualize HEALPix data, which takes advantage of the latest technology (SwiftUI and Metal libraries) and provides a number of improvements both in convenience and performance over existing solutions. It implements a strict superset of \texttt{map2gif} features (including original color palettes and projections), is fully GPU accelerated, and is geared towards interactive data analysis and exploration. While perhaps not as flexible as what could be achieved with \texttt{healpy} given enough effort, it is much faster and more convenient. HEALPix Viewer is capable of handling extremely large maps (such as those that will be produced by Simons Observatory) without any shortcuts or approximations, and produces high-quality output for publication or direct inclusion into presentations via drag and drop. Both \href{https://github.com/andrei-v-frolov/healpix-viewer}{source code} and \href{https://www.sfu.ca/physics/cosmology/healpix/}{binaries} are available for developers and researchers.

The paper is organized as follows: Section~\ref{sec:format} briefly describes HEALPix data format and pixelization of a sphere, Section~\ref{sec:pipeline} explains overall data processing pipeline structure, Section~\ref{sec:mapping} goes over color mapping and visualization strategy, Section~\ref{sec:analysis} focuses on data analysis tools provided, while Section~\ref{sec:interface} briefly goes over the user interface. Software dependencies are listed in Section~\ref{sec:dependencies}, testing procedures are described in Section~\ref{sec:testing}, performance and benchmarks are presented in Section~\ref{sec:performance}, while limitations are discussed in Section~\ref{sec:limitations}. Current development roadmap is outlined in Section~\ref{sec:development}, while conclusions are presented in Section~\ref{sec:conclusions}. Appendices go over spherical projections implemented (\ref{sec:projection}), Euler angles and generators of rotation (\ref{sec:representation}), dynamics of a sphere used for animation of viewpoint transitions (\ref{sec:integrator}), implementation of inverse error function used for map normalization (\ref{sec:erfinv}), and basic color science used for visualizations (\ref{sec:color}).

\section{HEALPix file format}
\label{sec:format}

HEALPix data is incapsulated in a \href{https://fits.gsfc.nasa.gov/}{FITS} (Flexible Image Transport System) file format, with specific metadata tags identifying the file as HEALPix data. FITS format is widely used in astronomy, and has many available I/O libraries targeting different languages or programming environments. As in original HEALPix code, we chose \href{https://heasarc.gsfc.nasa.gov/fitsio/}{\code{cfitsio} library}, developed and maintained by NASA's HEASARC. It is efficient and highly portable, and compiles out of the box on Intel and Arm64 architectures, which covers all of the Apple hardware (including iPhone). Accessing \href{https://developer.apple.com/documentation/swift/using-imported-c-functions-in-swift}{C functions from Swift} is easy once \href{https://developer.apple.com/documentation/swift/importing-objective-c-into-swift}{bridging header} is generated (which can be done automatically).

HEALPix metadata parser was written from scratch, using Swift facilities not available in Fortran (such as complex enumeration types and dictionaries), which simplified execution logic to a large extent. Complete specification of HEALPix format is available \href{https://healpix.sourceforge.io/data/examples/healpix_fits_specs.pdf}{here}, but we had to loosen the strict requirements on some tags to read the data as produced and published by major experiments. Absolutely necessary are \code{PIXTYPE = "HEALPIX"} tag identifying file as containing HEALPix data, and \code{NSIDE} and \code{ORDERING} tags specifying map resolution and ordering scheme (can be either \code{RING} or \code{NESTED}) of the map. \code{POLAR} tag indicates polarization data is available (as Stokes $Q$ and $U$ parameters), and has two sign conventions supported (\code{IAU} and \code{COSMO}). Map data itself is stored in a binary table, chunked and potentially having different data formats for different columns. The data reading code is ported from original Fortran implementation in HEALPix, using \code{cfitsio} routines called directly from Swift.

\begin{figure}
  \begin{center}
    \epsfig{file=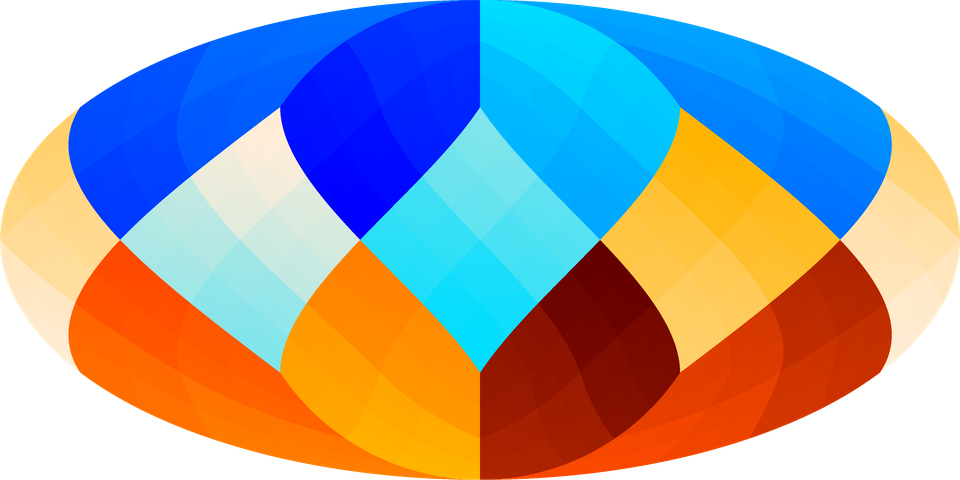, width=\columnwidth}
  \end{center}
  \caption{HEALPix pixelization of a sphere colored according to \code{NESTED} pixel order (in Mollweide projection). 12 faces are clearly distinguishable, and within each, pixels are indexed in interleaved bit pattern (known as Morton ordering).}
  \label{fig:nested}
\end{figure}

HEALPix pixelization of a sphere provides equal area and iso-latitude pixels, facilitating efficient spherical transforms using fast Fourier transform for $\theta=\text{const}$ rings, while using Legendre transforms in $\theta$ direction \citep{2005ApJ...622..759G, 2013A&A...554A.112R}. \code{RING} storage ordering enumerates pixels along iso-latitude rings and is best suited for spherical transform computation, while \code{NESTED} ordering enumerates pixels in a hierarchical fashion for 12 faces (using interleaved $x$ and $y$ bit pattern, also known as Morton ordering in image processing literature) and is best suited for localized memory access. To keep things simple, HEALPix Viewer converts \code{RING} data into \code{NESTED} ordering on input, and all further processing is done on \code{NESTED} maps. To illustrate the order in which \code{NESTED} pixels are enumerated, Figure~\ref{fig:nested} shows pixel index colored with Planck palette (deep blue to dark broun) in Mollweide projection, as seen from inside of the celestial sphere.

\section{Processing pipeline}
\label{sec:pipeline}

Primary design goal of HEALPix Viewer is fast and easy data visualization, so implementation decisions were made in favour of performance versus ultimate accuracy. Processing pipeline is 32-bit floats (64-bit float maps are down-converted at load time), and hardware acceleration with graphical processing unit (GPU) is used as much as possible. Basic 1-point statistic tools are provided for convenience (PDF, CDF, various moments of the distribution), but neither multi-point statistics nor spherical transforms are implemented (the latter due to the fact that there are no GPU-accelerated spherical transform libraries available so far).

\begin{figure}
  \begin{center}
  \begin{tikzpicture}[node distance=2cm]
    \tikzstyle{arrow} = [thick,->,>=stealth]
    \tikzstyle{input} = [rectangle, minimum width=1cm, minimum height=0.3cm,text centered, draw=black]
    \tikzstyle{buffer} = [trapezium, trapezium left angle=70, trapezium right angle=110, minimum width=1cm, minimum height=0.3cm, text centered, draw=black, fill=blue!30]
    \tikzstyle{func} = [diamond, minimum width=1cm, minimum height=0.3cm, text centered, draw=black, fill=green!30]

    \node (file) [input] {\texttt{.fits} file};
    \node (map) [buffer,below of=file,yshift=1cm] {\texttt{NESTED} map};
    \draw [arrow] (file) -- (map);
    \node (fx) [buffer,left of=map,xshift=-0.5cm] {$f(x)$ map};
    \node (cdf) [buffer,right of=map,xshift=0.5cm] {CDF map};
    \draw [arrow] (map) -- (fx);
    \draw [arrow] (map) -- (cdf);
    \node (color) [func,below of=map] {color mapper};
    \node (palette) [buffer,left of=color,xshift=-0.6cm] {palette};
    \node (colormap) [input,below of=palette,yshift=1.0cm] {color map};
    \draw [arrow] (map) -- (color);
    \draw [arrow] (fx) -- (color);
    \draw [arrow] (cdf) -- (color);
    \draw [arrow] (palette) -- (color);
    \draw [arrow] (colormap) -- (palette);
    \node (texture) [buffer,below of=color] {face textures};
    \draw [arrow] (color) -- (texture);
    \node (mapper) [func,below of=texture,yshift=0.4cm] {mapper};
    \node (projection) [input,left of=mapper,xshift=-0.6cm] {geometry};
    \node (lighting) [input,right of=mapper,xshift=0.6cm] {lighting};
    \draw [arrow] (texture) -- (mapper);
    \draw [arrow] (projection) -- (mapper);
    \draw [arrow] (lighting) -- (mapper);
    \node (drawable) [buffer,below of=mapper,yshift=0.4cm] {drawable};
    \draw [arrow] (mapper) -- (drawable);
    \node (transform) [input,above of=fx,yshift=-1cm] {transform};
    \draw [arrow] (transform) -- (fx);
    \node (annotation) [input,left of=drawable,xshift=-0.5cm] {annotation};
    \draw [arrow] (annotation) -- (drawable);
  \end{tikzpicture}
  \end{center}
  \caption{Schematic of the data flow in HEALPix Viewer. Rectangles represent user input, blue-filled trapezoids represent GPU buffers and textures, green-filled diamonds represent GPU kernels. Lazy evaluation is used in color mapper, which is the main workload for high resolution maps. Geometry mapper runs at output resolution, and is cheap.}
  \label{fig:flow}
\end{figure}
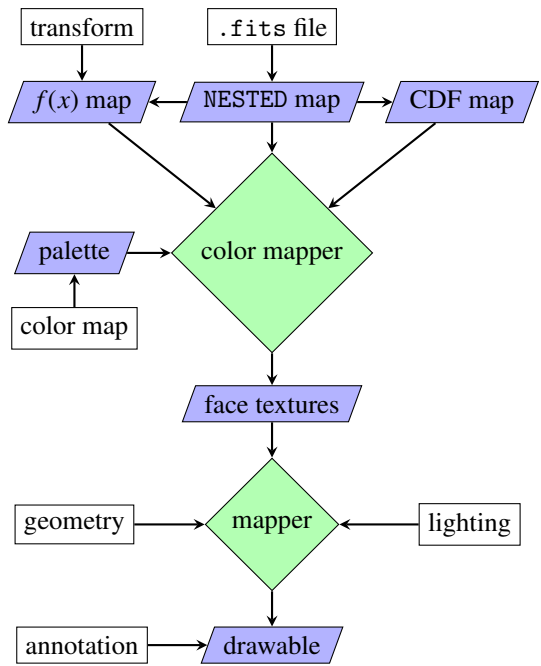

Overall structure of the data flow and user inputs is illustrated in Figure~\ref{fig:flow}. HEALPix data gets loaded from disk into CPU memory, down-converted to 32-bit floats and rearranged into \code{NESTED} ordering if necessary, and then loaded to GPU buffer (on shared memory architecture the last step does not involve data transfer). A functional transform based on user input is optionally applied via GPU kernels, and transformed map is stored into another GPU buffer of the same size as original map. Meanwhile, map data is dispatched for sorting and ranking in background for construction of the CDF, ranked statistics, and equalized and normalized maps. As sorting algorithms are generally hard to parallelize efficiently, this step is carried out by CPU, albeit ranking of all of the loaded maps is dispatched concurrently to exploit CPU parallelism to the extent possible. Once constructed, ranked map is loaded into another GPU buffer.

Once the data is loaded into GPU memory, data values can be mapped into full color representation suitable for rendering. Color mapper uses internal HEALPix representation of a sphere as 12 faces covered by square bitmaps as output, producing a 2D texture array. Texture array (as opposed to array of textures) is an atomic data type supported on many GPU architectures, and could employ advanced memory management techniques such as hardware swizzling to improve performance, with calling functions being isolated from particular details by samplers. Color mapping is also hardware accelerated, with user-specified color palette preloaded as 1D texture, and hardware linear interpolation sampler used to map data values into pixel color. Dispatch of the color mapper kernel is done as a 3D wavefront in $x$, $y$, and face index. As a result, the data buffer access pattern tends to be localized because, in \code{NESTED} ordering, pixels are stored sequentially by face in a bit-interleaved x, y order, as in Morton ordering. No particular effort has been put into color management, with RGBA pixel values interpreted in device colorspace (usually sRGB), which is the convention established by \code{map2gif} in the original HEALPix code.

\begin{figure}
  \begin{center}
  \begin{tabular}{cc}
    Mollweide & Hammer \\
    \epsfig{file=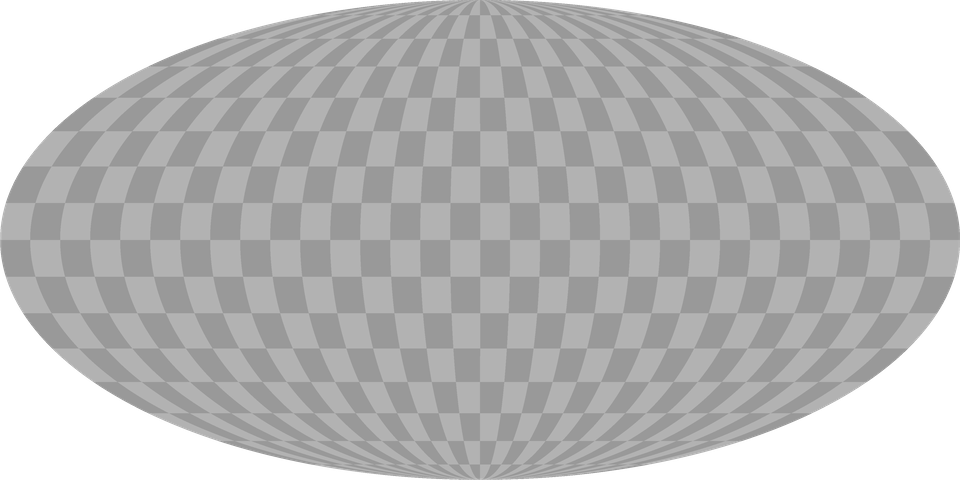, width=4cm} &
    \epsfig{file=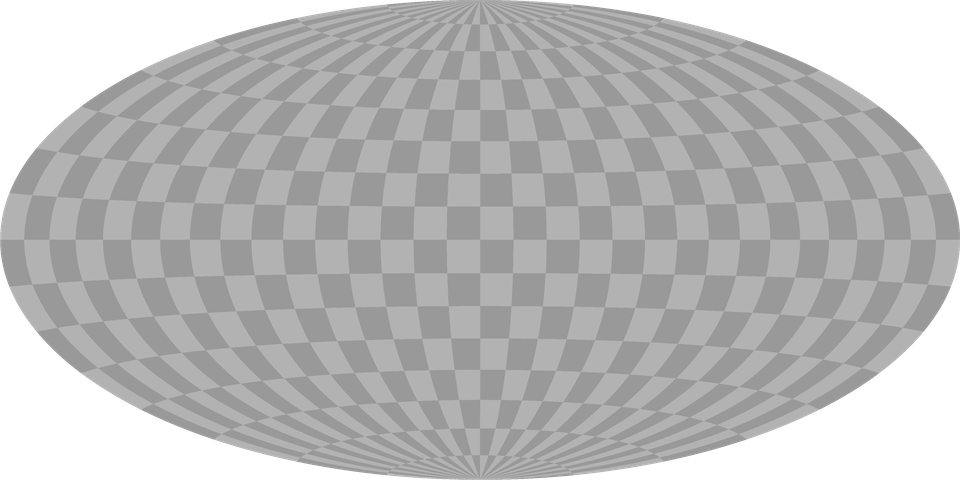, width=4cm} \\
    Lambert & Isometric \\
    \epsfig{file=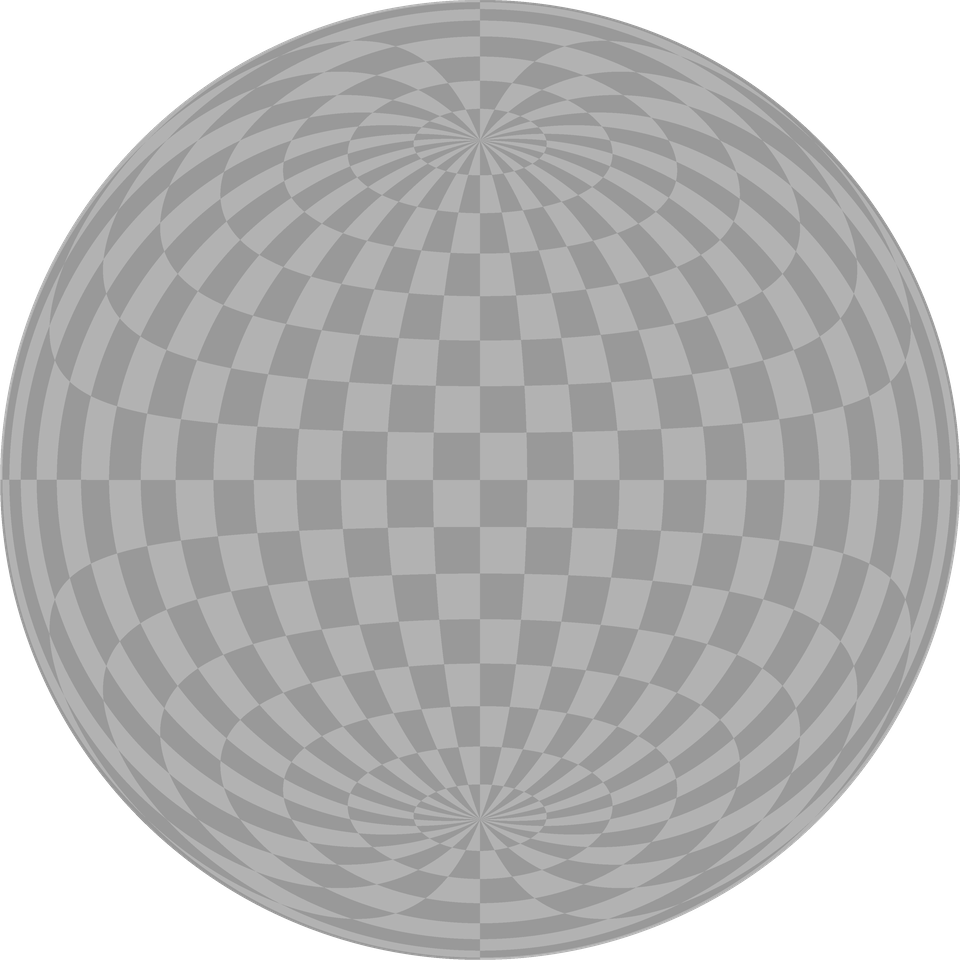, width=4cm} &
    \epsfig{file=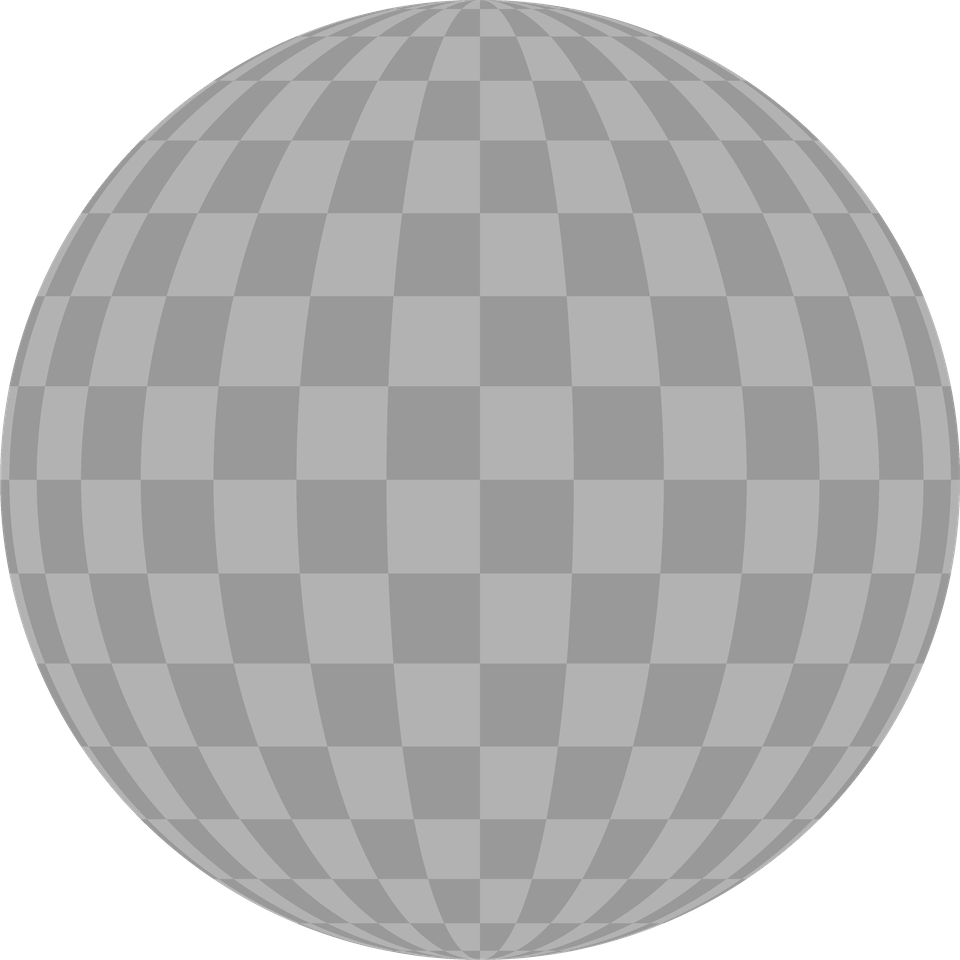, width=4cm} \\
    Gnomonic & Werner \\
    \epsfig{file=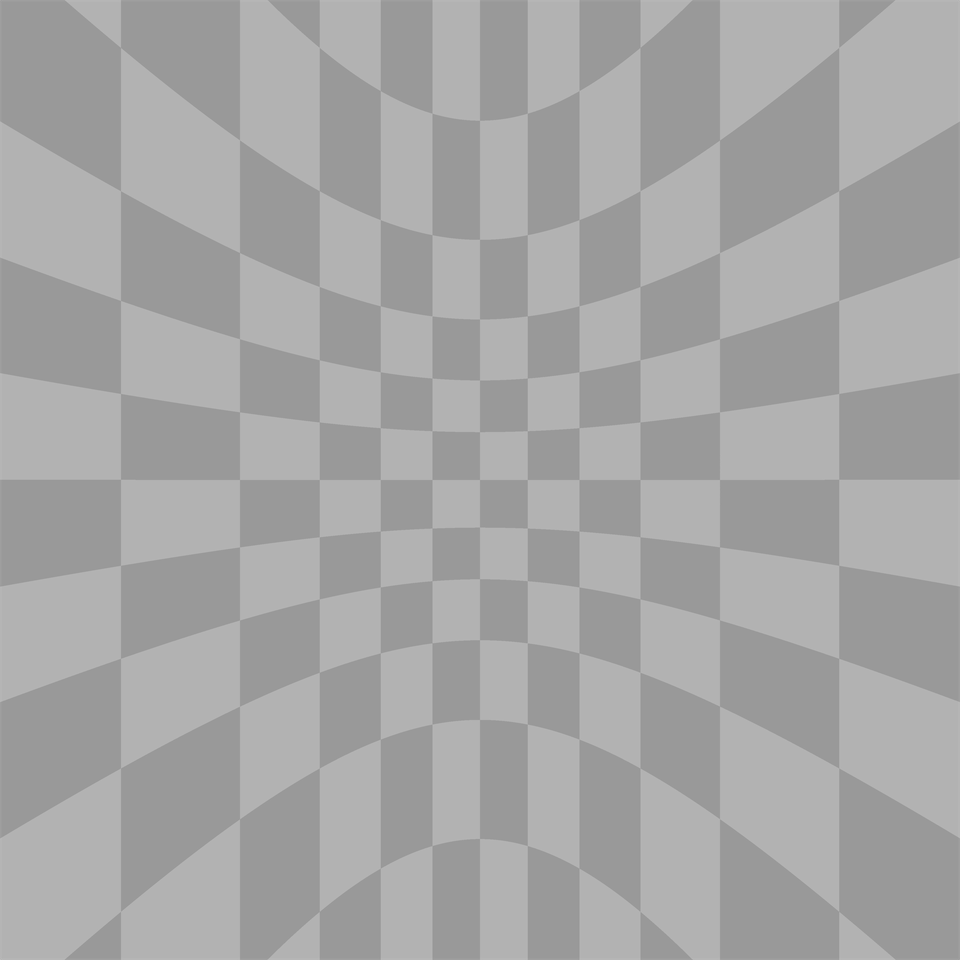, width=4cm} &
    \epsfig{file=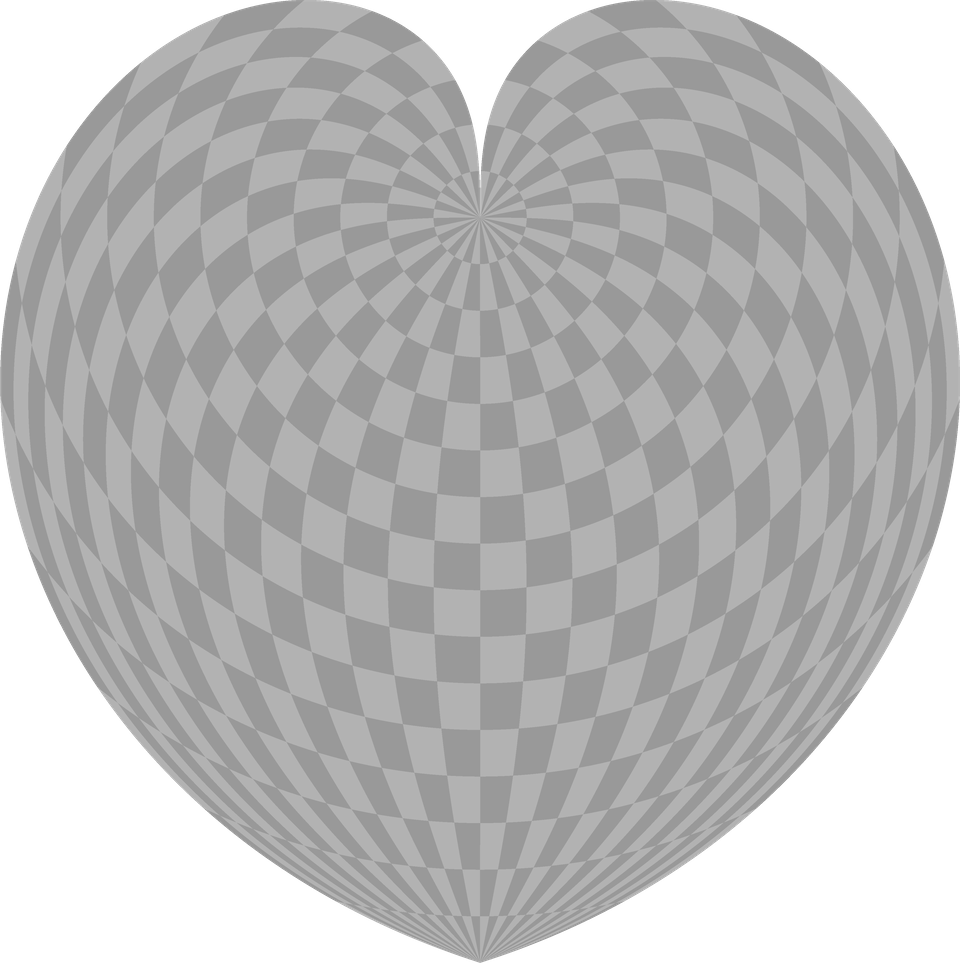, width=4cm} \\
  \end{tabular}
  \end{center}
  \caption{Some of the projections available in HEALPix Viewer. Checker-board pattern corresponds to the lines of constant latitude and longitude, as projected onto a two-dimensional screen plane from equator and zero longitude viewpoint.}
  \label{fig:projection}
\end{figure}

All of the above is done via lazy evaluation, i.e.\ the data is processed only when parameters change, and operates at full map resolution. In contrast, user-facing view is updated at hardware monitor refresh rate (usually 60fps on most modern Macs) to present smooth animations, with rendering done synchronously and directly into the drawable presented to the user. This is carried out at the screen resolution, and would effectively be down-sampling of the data for Planck ($\text{nside}=2048$) and especially Simons Observatory ($\text{nside}=8192$) maps. A down-sampling strategy always involves a compromise between what is kept versus what is filtered out. In HEALPix viewer, nearest neighbour sampling was chosen as a main strategy for the reason that it preserves 1-point PDF of the high-resolution data, although other levels of data averaging are supported in hardware, with two being available as customized presets in the current version. Oversampling of the output for export is optionally available (which produces visually more pleasing maps), and is implemented via Lanczos interpolation, once again in hardware.

As all the user-facing images have planar geometry, projection of a sphere must be done, which is carried out by the geometry mapper. HEALPix Viewer supports many projections out of the box, some of which are illustrated in Figure~\ref{fig:projection}. For complete list and projection formulae, refer to \ref{sec:projection}. Implementing a new one is a matter of a few lines of code, due to modular design. The screen coordinates $(X,Y)$ get converted to cartesian plane ones $(x,y)$ via affine transformation
\begin{equation}
  \left[ \begin{array}{c} x\\ y \end{array} \right] =
  \mathbb{M} \cdot \left[ \begin{array}{c} X\\ Y \\ 1\end{array} \right],
\end{equation}
where transform $\mathbb{M}$ is a 2x3 matrix precomputed on CPU side and passed as a constant parameter to geometry mapper kernels. It depends on screen resolution, viewport size, zoom level, and desired output alignment. It can also incorporate coordinate flips (screen $Y$ direction is usually upward, while bitmap $Y$ direction is downward), viewpoint inversion (looking at the sphere from inside versus outside effectively flips $x$ direction), and overall rotation of the output image (currently not used). The plane coordinates are then mapped in the geometry mapper kernel to a unit 3D vector $\vec{u}$ via inverse spherical projection $\vec{u} = p(x,y)$, and if the supplied values are outside of bounds of projected sphere, background color is used for output.

As one would often like to rotate the sphere to view a particular location, a rotation matrix $\mathbb{R}$ is applied to vector $\vec{u}$, with
\begin{equation}
  \vec{v} = \mathbb{R}\cdot\vec{u}
\end{equation}
finally used to lookup the pixel color in 2D texture array produced by color mapper. Optionally, lighting effects could be applied, with overall pixel brightness depending on light direction $\vec{l}$ via scalar product $\vec{l}\cdot\vec{u}$ (light location is fixed in the viewer space and is not rotated with the map).

\begin{figure}
  \begin{center}
    \epsfig{file=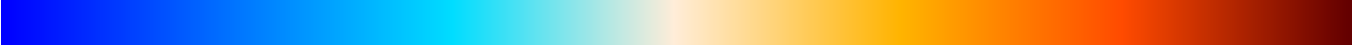, width=\columnwidth}\\
    \epsfig{file=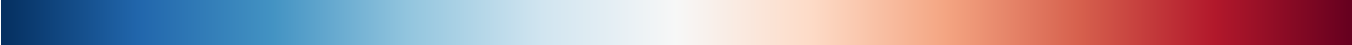, width=\columnwidth}\\
    \epsfig{file=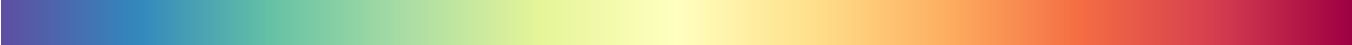, width=\columnwidth}\\
    \epsfig{file=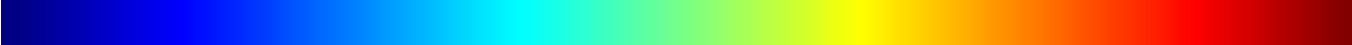, width=\columnwidth}\\
    \epsfig{file=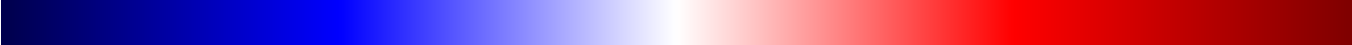, width=\columnwidth}\\
    \epsfig{file=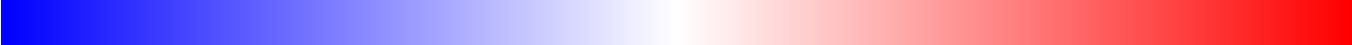, width=\columnwidth}\\
    \epsfig{file=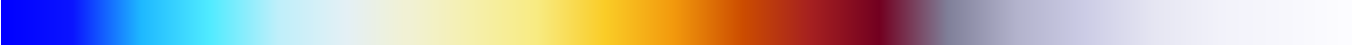, width=\columnwidth}\\
    \epsfig{file=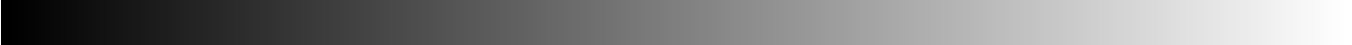, width=\columnwidth}\\
    \epsfig{file=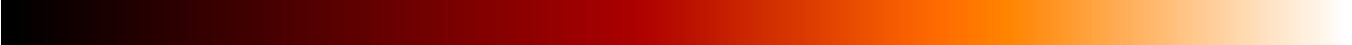, width=\columnwidth}\\
    \epsfig{file=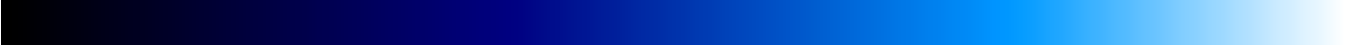, width=\columnwidth}\\
    \epsfig{file=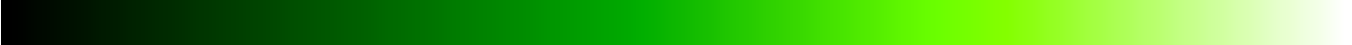, width=\columnwidth}\\
    \epsfig{file=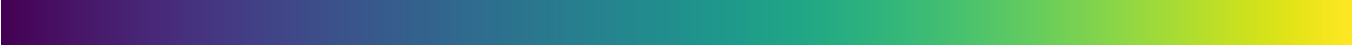, width=\columnwidth}\\
    \epsfig{file=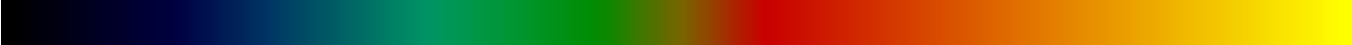, width=\columnwidth}\\
    \epsfig{file=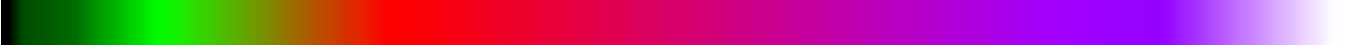, width=\columnwidth}
  \end{center}
  \caption{Colormaps hard-coded in HEALPix Viewer include a superset of original HEALPix colormaps, colormaps used by Planck, and a few commonly used ones from Python. From top to bottom: Planck (aka.\ Parchment), Faded (inverted RdBy), Spectral (Python), original HEALPix CMB, Seismic (Python), Difference (Python), Frequency map (Planck), Greyscale (HEALPix), Hot (HEALPix), Cold (HEALPix), Lime (channel swap of Hot/Cold), Viridis (Python), BGRY (HEALPix), GRV (HEALPix).}
  \label{fig:gradients}
\end{figure}

Rotation matrix $\mathbb{M}$ is usually parametrized by Euler angles, which are latitude, longitude, and azimuth of the viewpoint (latitude and longitude can be specified in \code{map2gif}, but not azimuth). Details of this rotation matrix parametrization, along with the way to extract Euler angles from rotation matrix are given in \ref{sec:rotation}. One of the features I wanted to implement is a smooth animation between two different orientations when changing the view. This is accomplished via evolution of generator of rotation as a damped simple harmonic oscillator. Rotation matrix in three dimensions can be represented as
\begin{equation}
  \mathbb{R} = \exp[\mathbb{W}],
\end{equation}
with antisymmetric 3x3 matrix $\mathbb{W}$ corresponding to a dual 3-vector $\vec{w} = *\mathbb{W}$, known as generator of rotation. This representation is quite familiar, as it appears in rotation of a solid body as angular velocity $\vec{\omega}$. Details of this representation are given in \ref{sec:generator}, along with formulae for forward and inverse transformations.  When user specifies a new target orientation $\vec{t}$, the current viewpoint generator $\vec{w}$ is evolved as
\begin{equation}
  \ddot{\vec{w}} + \gamma \dot{\vec{w}} + \kappa(\vec{w} - \vec{t}) = 0,
\end{equation}
with rotation matrix $\mathbb{R}$ updated correspondingly for each frame, using real system time in case frame cadence is not uniform. Parameters $\gamma$ and $\kappa$ set timescales of damping and oscillation correspondingly, with the values chosen so that evolution is slightly overdamped with timescale of a few seconds. Evolution itself is computed with 6-th order symplectic integration scheme \citep{1990PhLA..150..262Y, 1991JCoPh..92..230C}, vectorized on CPU side, which is not strictly necessary but is easy to implement and fast. Details of this are presented in \ref{sec:integrator}.

Finally, the rendered image might be saved into a file for use with other software or as paper figures. PNG format is chosen as a default because it supports full 24-bit colour and transparency, in which it is superior to GIF format used by \code{map2gif}. Additionally, GIF, HEIF, and 16-bit TIFF image formats are available for export. Formats based on lossy Fourier compression such as JPEG are not suitable for high-quality output due to compression artifacts. Rendering for export is done by the same GPU code as for the screen, except affine transformation parameters are adjusted accordingly, and rendering is done into 32-bit or 64-bit integer RGBA context. Colorbar render can be added to the figure if desired, using the same strategy. One might also wish to have a text annotation, which is rendered using Core Text framework, and can use any fonts available on the system (San Francisco Compact is used by default). Resulting texture is copied into CPU bitmap, which is output as an image file using Core Image routines.

\section{Color mapping}
\label{sec:mapping}

\begin{figure}
  \begin{center}
    \epsfig{file=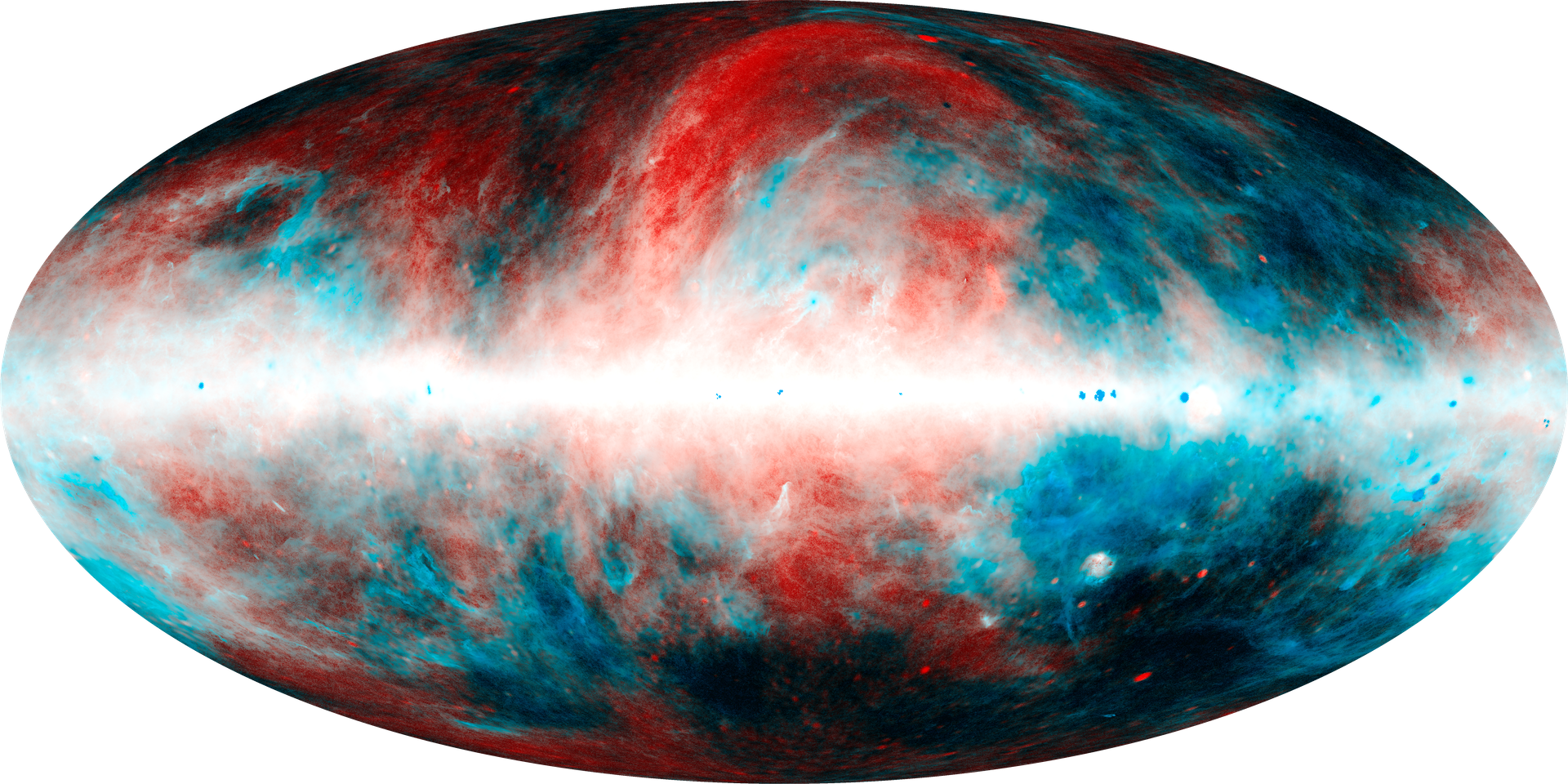, width=\columnwidth}\bigskip\\
    \epsfig{file=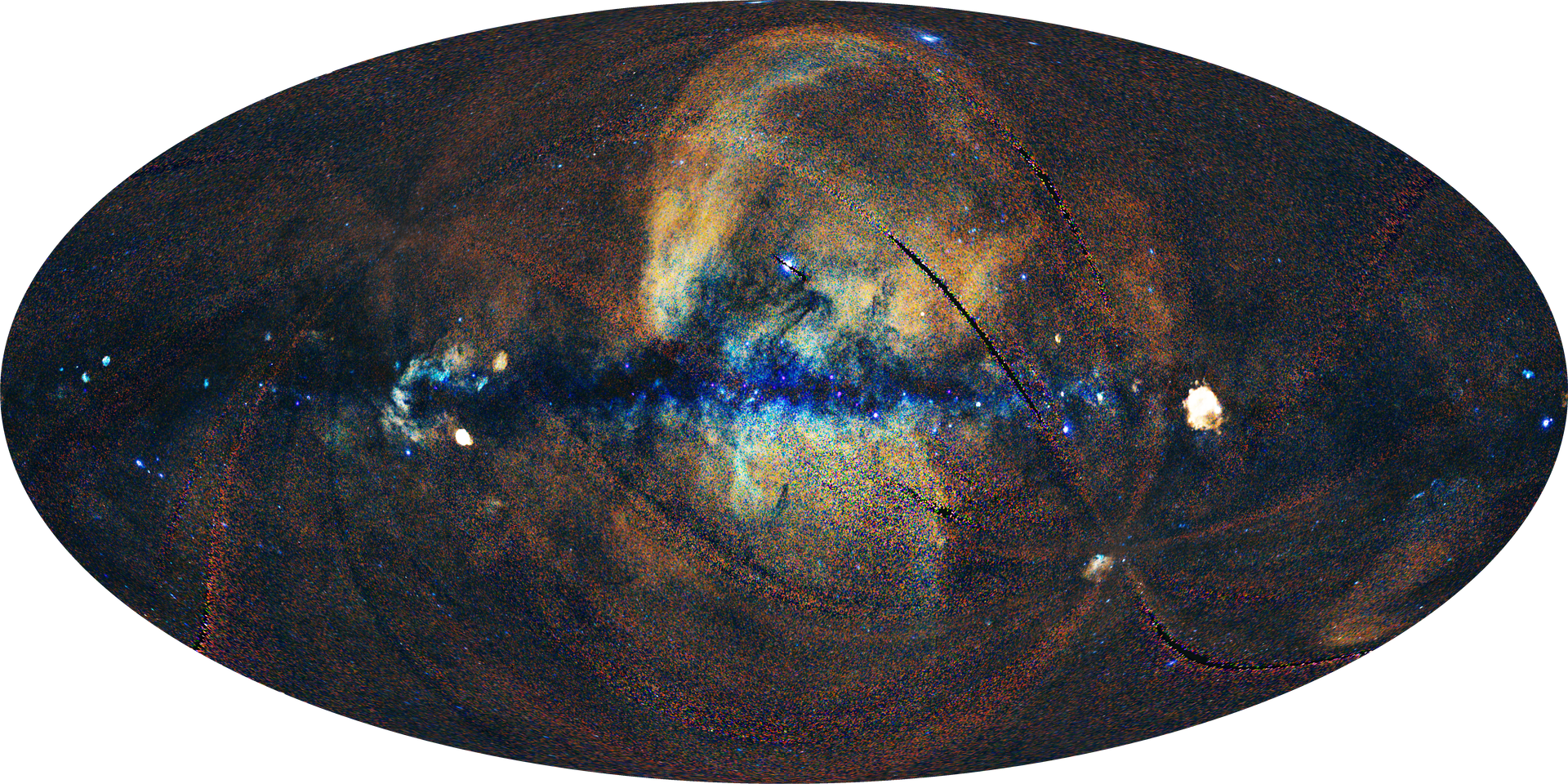, width=\columnwidth}
  \end{center}
  \caption{False color maps generated by HEALPix Viewer, presented in Mollweide projection. (a) Planck synchrotron and dust emission with synchrotron component in red (upscaled Commander 2015 data) and dust component in cyan (CMB-subtracted 353GHz Planck 2018 map). Both components are equalized prior to mixing. (b) ROSAT R4, R5, and R6 X-ray bands decorrelated and presented in false color.}
  \label{fig:falsecolor}
\end{figure}

Data values of a map to be visualized are typically represented by colors drawn from a specified palette, with data values from some interval $[a,b]$ mapped into a unit range
\begin{equation}
  x' = (x-a)/(b-a),
\end{equation}
and user-selectable colormap applied. HEALPix Viewer implements a number of colormaps out-of-the-box, in particular original HEALPix and Planck colormaps, along with a few commonly used ones from Python, as illustrated in Fig.~\ref{fig:gradients}. Values below or above specified range are clipped, and are represented by user-defined min and max colors (defaulting to colormap limits). In addition, color of non-representable (\texttt{NaN}) values and background can be specified. Transparency is fully supported.

In the current development version, custom colormaps interpolating between a list of user-specified colors are implemented (blending in perceptually uniform color space, as described in \ref{sec:oklab}), with proper GUI soon to be added.

Among the more advanced visualization capabilities of the HEALPix Viewer is the ability to generate false color maps to represent multi-band data sets. While it is fairly common in astronomy or high-energy astrophysics, it is under-utilized in CMB literature, possibly due to the fact that neither \texttt{map2gif} nor \texttt{healpy} support it natively. HEALPix Viewer does support false color maps, along with a few options to decorrelate the data on the fly. The utility of this approach is obvious when looking at Figure~\ref{fig:falsecolor}, where correlation of synchrotron component with X-ray emission is visually evident, while dust component is clearly responsible for absorption of X-rays. Planck foreground data \cite{2016A&A...594A..25P} was previously plotted against that of ROSAT \citep{1997ApJ...485..125S}, with eROSITA DR1 release \citep{2021A&A...647A...1P} slated to provide much-improved data soon, opening exciting possibilities. 

A common problem in visualizing multi-band data is that of channels being highly correlated, which leads to color dimension being under-utilized by simple linear combination. HEALPix viewer implements parametric whitening, where channel data $\{X_i\}$ can be continuously scaled from original with covariance $V=\text{Cov}(X_i,X_j)$ and correlation $P=\text{Corr}(X_i,X_j)$, to whitened $Y = V^{-\frac{1}{2}} X$ and maximally decorrelated $Z = P^{-\frac{1}{2}} Y$. Matrix inversion is done in SVD sense, with eigenvalues of $V^{-\frac{1}{2}}$ regularized as
\begin{equation}
  \sigma_i = \frac{\lambda_i}{\left(\epsilon^2 + \lambda_i^2\right)^{\frac{3}{4}}}, \hspace{1em}
  \epsilon = 10^{-4} \max \{\lambda_i\},
\end{equation}
accommodating perfectly correlated input maps. Decorrelated data is used as coefficients of red, green, and blue primaries (specified by the user and optionally scaled to a particular white point), with gamut compression optionally applied as described in \ref{sec:gamut}.

\section{Data analysis}
\label{sec:analysis}

\begin{figure}
  \begin{center}
    \epsfig{file=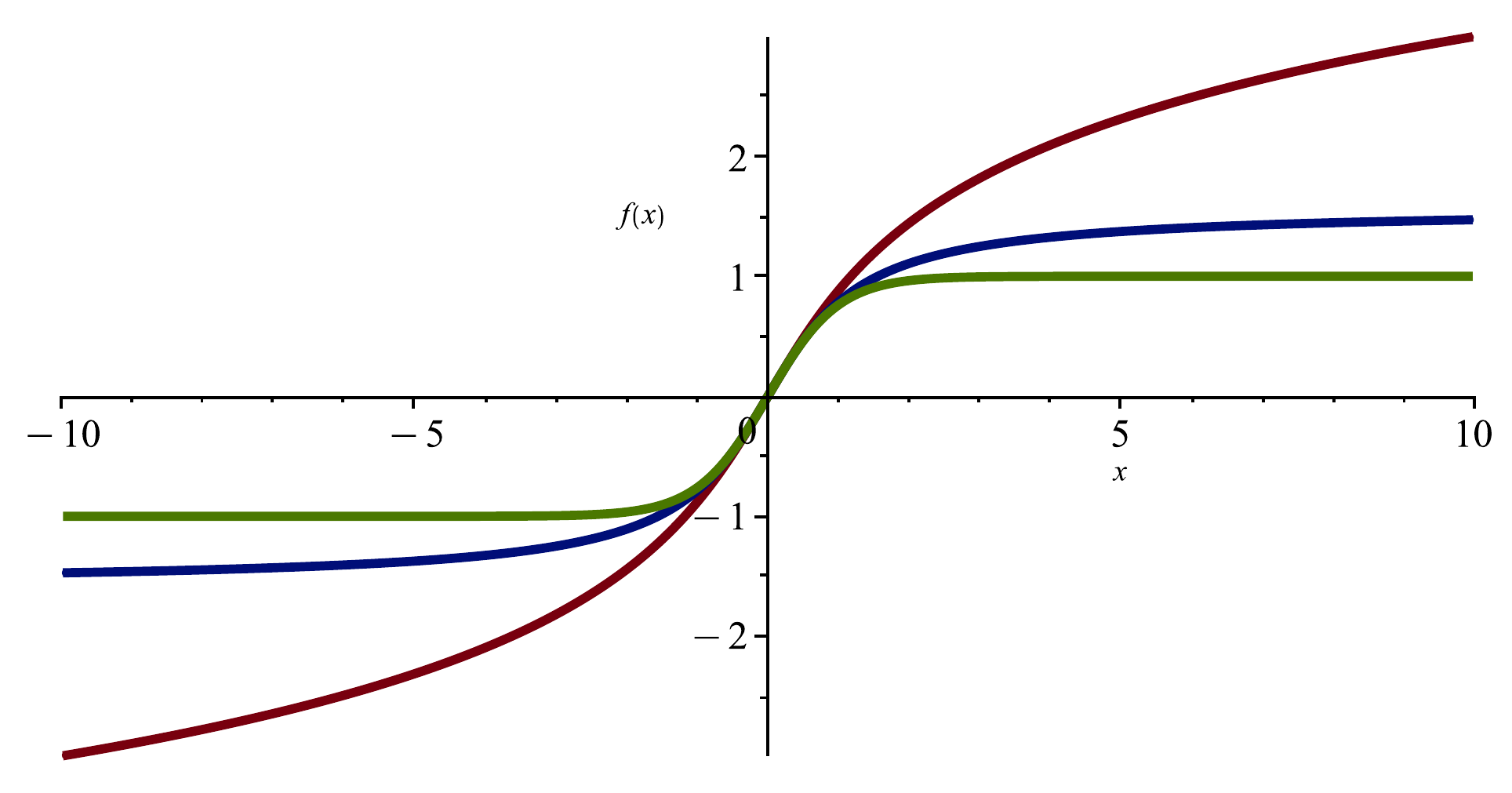, width=\columnwidth}
  \end{center}
  \caption{Three range-limiting transforms used in HEALPix Viewer. Dark brown line is asinh transform, dark blue is atan transfrom, and dark green is tanh transfrom. All three have the same derivative near the origin, but differ in asymtotics for large argument values (logarithmic, rational, and exponential fall-off correspondingly).}
  \label{fig:transform}
\end{figure}

\begin{figure}
  \begin{center}
    \epsfig{file=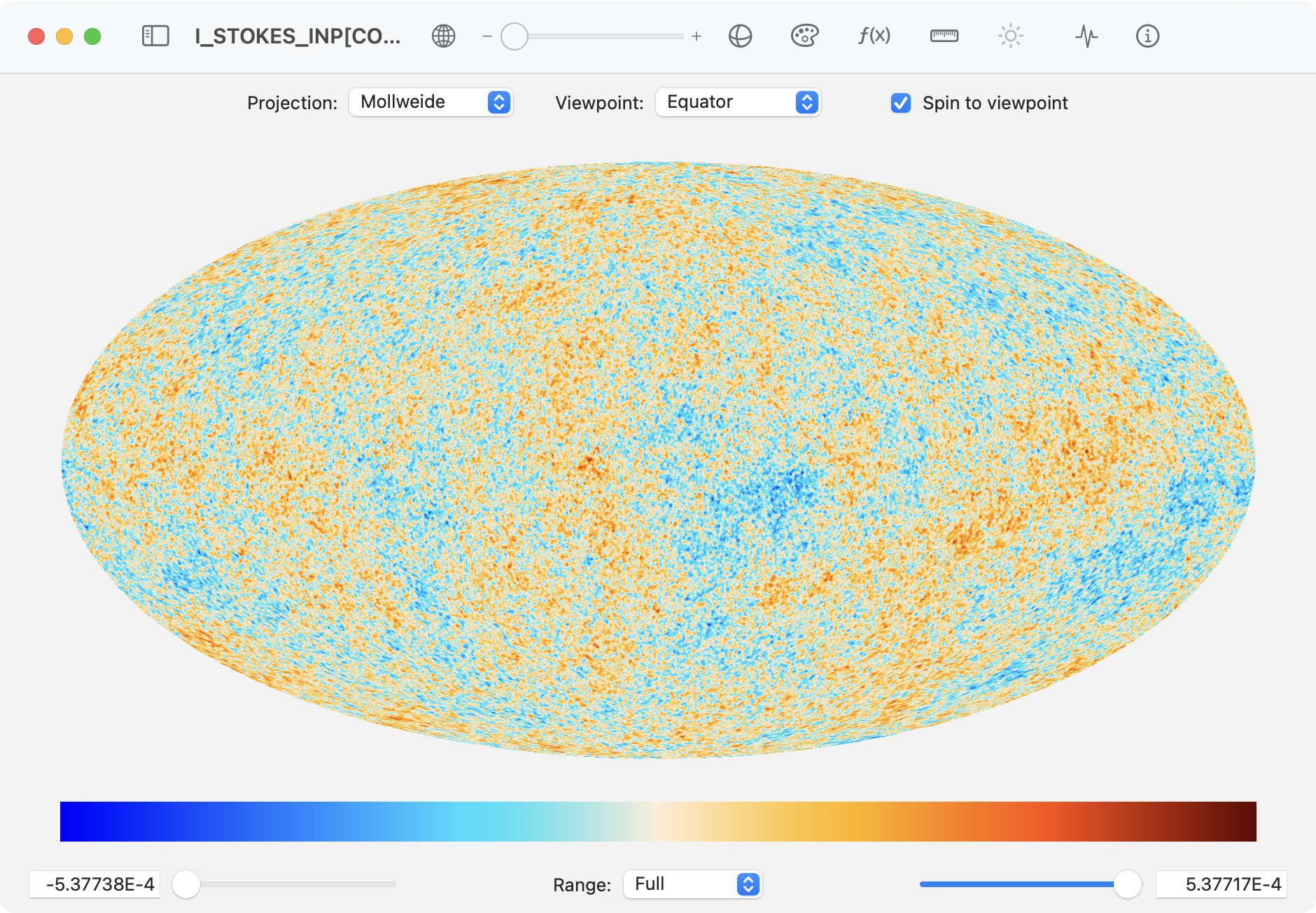, width=\columnwidth}
  \end{center}
  \caption{HEALPix Viewer user interface displaying inpainted Planck 2018 SMICA CMB map, with top and bottom toolbars exposed.}
  \label{fig:ui:main}
\end{figure}

While sophisticated data analysis is not the primary goal for HEALPix Viewer and is best carried out by dedicated code, some data analysis facilities are provided for convenience and improved visualization. These are currently restricted to 1-point transformations, where a pixel value $x$ gets mapped into a value of a function $f(x)$. The list of implemented transformations is as follows:
\begin{itemize}
  \setlength\itemsep{0em}
  \item $f(x) = \ln(x-\mu)$
  \item $f(x) = \text{asinh}[(x-\mu)/\sigma]$
  \item $f(x) = \text{atan}[(x-\mu)/\sigma]$
  \item $f(x) = \tanh[(x-\mu)/\sigma]$
  \item $f(x) = \pm |x-\mu|^\sigma$
  \item $f(x) = \exp[(x-\mu)/\sigma]$
  \item PDF equalization
  \item PDF normalization
\end{itemize}
Parameter $\mu$ corresponds to the nominal center of the distribution, while parameter $\sigma$ corresponds to overall distribution width, and is set on logarithmic scale in the user interface. Logarithmic scaling (first transform on the list) is widely used to compress dynamical range of data for visualization, and is in particular useful for representing intensity of the higher frequency maps in CMB experiments (such as 353GHz, 545GHz, and 857GHz data in Planck) as foregrounds in galactic plane are vastly brighter than high-latitude diffuse emission. Logarithmic scaling has a hard lower cut-off, with data $x\le\mu$ resulting in non-normalizable output (colored as \code{NaN} color in HEALPix Viewer).

The next three transforms correspond to symmetric and increasingly harder clamping of input data, as illustrated in Figure~\ref{fig:transform}. Power law and exponential transforms could be used to expand the data instead. The last two transforms deserve a special mention, as they are rarely if ever seen in CMB literature, but are common in image processing and are very useful for data visualization. PDF equalization applies a function which produces output with uniform 1-point PDF, while normalization extends on that by producing Gaussian 1-point PDF. By themselves, they are useful pre-processing steps to apply to data for multi-variate and multi-point analysis (c.f.\ a notion of copula in statistics), and they also serve for producing almost optimal color gradation in color-mapped image, providing a one-button ``make it pretty'' option.

CDF estimators and ranked maps are produced by indexing the map data by its value, which is done using \href{https://github.com/scandum/quadsort}{quadsort} algorithm, slightly modified to be thread-safe. While not as widely known as quicksort, quadsort is stable (preserves ordering of equal values), on average a bit faster, and scales vastly better ($O(n\log n)$ versus $O(n^2)$) for the worst case (ordered data, which appears all the time in masked maps). CDF estimator is then easily computed as
\begin{equation}
F(x) = \{ \text{\# of pixels with value} < x\}/n
\end{equation}
while output map with uniform PDF is achieved by transforming the ranked map pixel $p$ into
\begin{equation}
f(p) = \{\text{rank of pixel } p\}/n.
\end{equation}
Current implementation has a side effect of having uniform value areas ranked in pixel order (since quadsort is stable), but fixing that is more trouble than it's worth (for CMB data it would only affect masked areas). Gaussian PDF can be produced by applying inverse Gaussian CDF transformation to uniform PDF map
\begin{equation}
f(x) = \sqrt{2}\, \text{erf}^{-1}(2x-1),
\end{equation}
which requires a bit of extra work to implement as $\text{erf}$ and its inverse are not available in either GPU nor CPU libraries \citep{erfinv}. Details of that are explained in \ref{sec:erfinv}.

\section{User interface}
\label{sec:interface}

\begin{figure}
  \begin{center}
    \epsfig{file=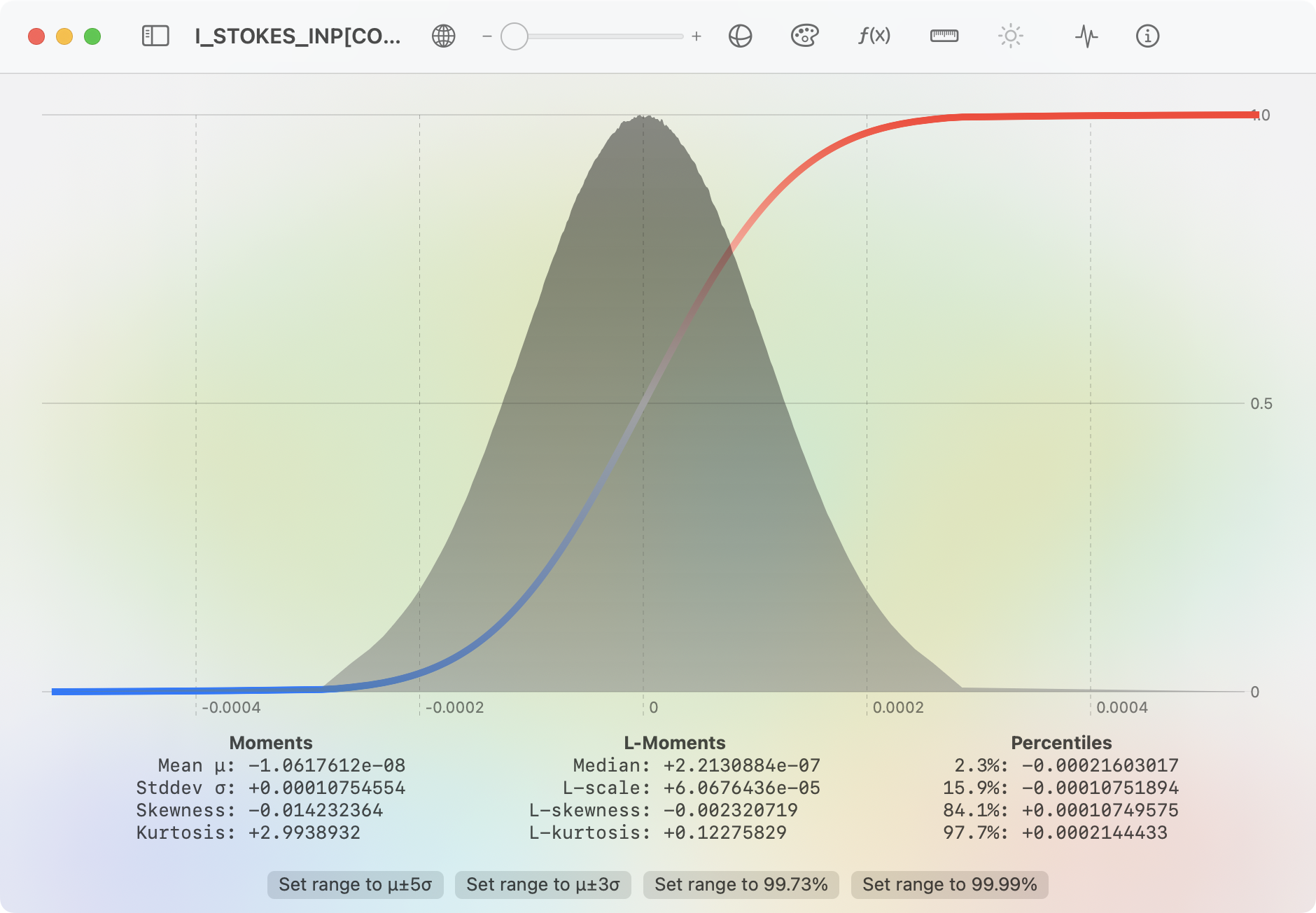, width=\columnwidth}
  \end{center}
  \caption{HEALPix Viewer statistics overlay displaying statistics summary of inpainted Planck 2018 SMICA CMB map.}
  \label{fig:ui:stats}
\end{figure}

User interface paradigm is based around top toolbar which allows user to adjust render parameters, and bottom toolbar which displays the colorbar and adjusts color mapping parameters. Top toolbar is hierarchical, and has several parameter groups which can be selected by buttons in the main window title bar. The top toolbar parameter hierarchy is as follows:
\begin{itemize}
  \setlength\itemsep{0em}
  \item map projection:
    \begin{itemize}
      \setlength\itemsep{0em}
      \item projection choice
      \item viewpoint presets
      \item inside/outside view
    \end{itemize}
  \item view orientation:
    \begin{itemize}
      \setlength\itemsep{0em}
      \item viewpoint latitude
      \item viewpoint longitude
      \item viewpoint azimuth
    \end{itemize}
  \item color scheme:
    \begin{itemize}
      \setlength\itemsep{0em}
      \item color palette
      \item below min color
      \item above max color
      \item \code{NaN} color
      \item background color
    \end{itemize}
  \item data transform:
    \begin{itemize}
      \setlength\itemsep{0em}
      \item function $f(x)$
      \item parameter $\mu$
      \item parameter $\ln\sigma$
    \end{itemize}
  \item lighting effects:
    \begin{itemize}
      \setlength\itemsep{0em}
      \item light latitude
      \item light longitude
      \item effect amount
    \end{itemize}
\end{itemize}
Screenshot of the user interface with both top and bottom toolbars exposed is shown in Figure~\ref{fig:ui:main}. If you're tight on screen space, you can hide both toolbars. Lightning effects have to be enabled in View menu if desired. Also in View menu are options to control application appearance (light or dark mode), and to enable cursor readout. Bottom toolbar shows the colorbar and the controls for its bounds. Colorbar range can be restricted to symmetric, positive, or negative if desired. Left sidebar (which also can be hidden, and is hidden by default) shows the list of loaded map data, and allows you to chose which one to display. More than one window can be opened simultaneously, with independent settings defaulting to the ones in Data menu. If Data menu selection changes, it will be applied to the active window.

In addition to toolbars, there are two information overlays available - data statistics and FITS header. They can be accessed by buttons in the title bar, and will temporary cover map view with a semi-transparent background. Data statistics overlay displays PDF and CDF (discontinuities of which are represented as a bar graph), as well as moments of the distribution (mean, standard deviation, skewness and kurtosis), L-moments \citep{Hosking:1990} (median, L-scale, L-skewness, and L-kurtosis), and percentile brackets of the distribution corresponding to $1\sigma$ and $2\sigma$ of Gaussian one, as shown in Figure~\ref{fig:ui:stats}. Buttons to quickly set colorbar range to preset values are added for convenience. FITS header overlay displays complete header of the FITS file, with all the cards and comments unaltered and unparsed. This is useful for debugging, as well as if one wants to check metadata not supported by HEALPix Viewer natively.

Main map view responds to common gestures - two finger pinch to zoom, two finger rotation to change azimuth, right click to center on a clicked location (rotating along a geodesic), and option-right click (rotating while keeping azimuth fixed). If animations are enabled, two-finger scroll gesture will kick the sphere in the direction of the scroll, while keeping target location which it will eventually return to. If cursor readout is enabled in the View menu, the coordinates and map value under the cursor will be displayed at the top of the map view.

Drag and drop is supported to enable simple integration with other applications, such as Powerpoint or Keynote for presentations. Drop HEALPix file into main window to load it, and drag the map and colorbar (individually) to the application which accepts image files as a drop. Dragged map will be exported at current screen size, with other options available in Settings dialog. If you want custom resolution output or oversampling, export the image via File menu.

\section{Software dependencies}
\label{sec:dependencies}

\begin{figure}
  \begin{center}
    \epsfig{file=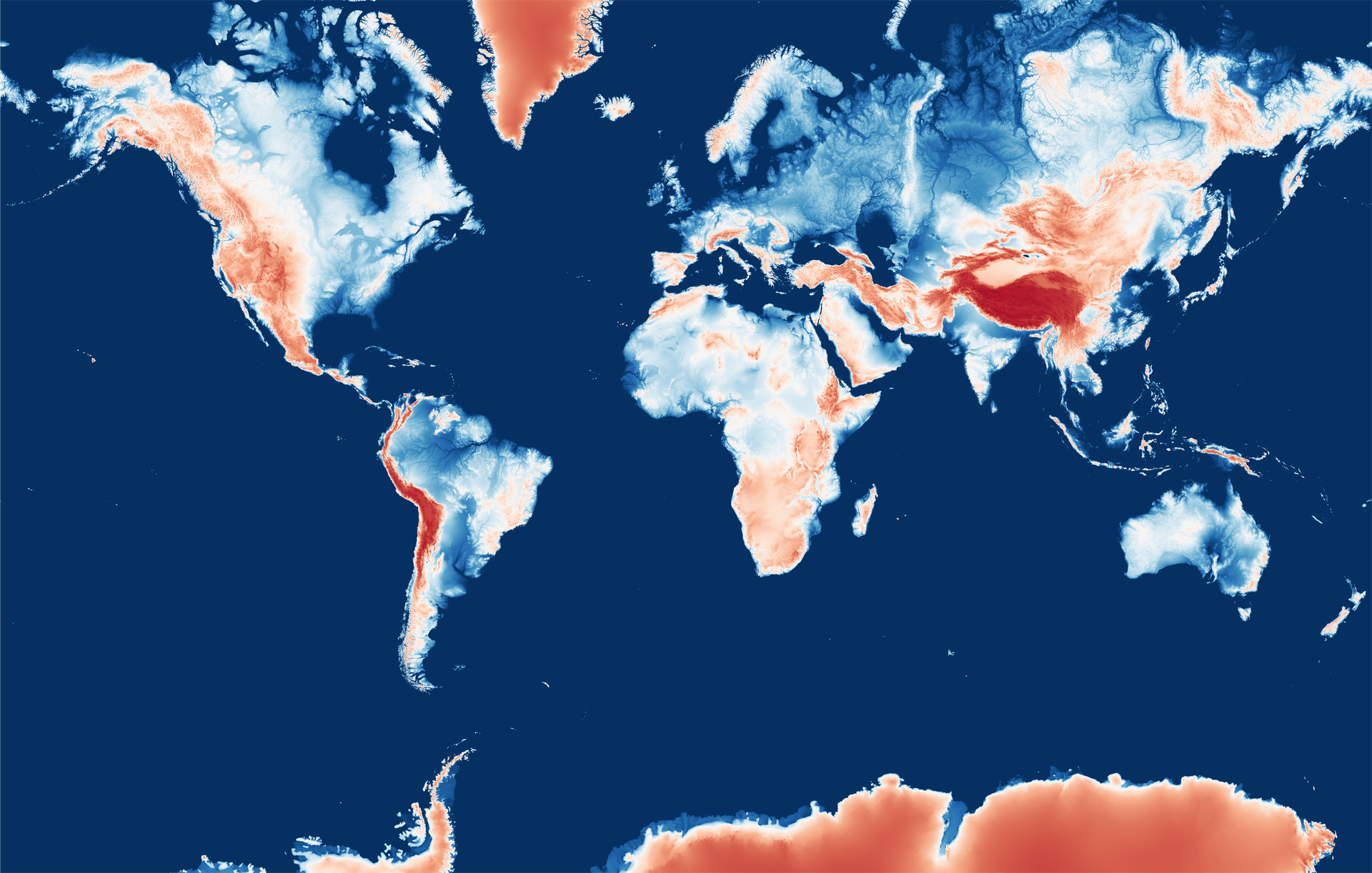, width=\columnwidth}
  \end{center}
  \caption{High-resolution ($n_{\text{side}}=8192$) digital elevation map of Earth used for testing and benchmarks, plotted in Mercator projection.}
  \label{fig:earth}
\end{figure}

HEALPix Viewer code is mostly self-contained, and depends only on a few external libraries for its functionality. To compile the current development source code, one will need
\begin{itemize}
  \setlength\itemsep{0em}
  \item macOS SDK v12.3+; v13+ is required for some features
  \item \href{https://apps.apple.com/ca/app/xcode/id497799835}{XCode compilers}; code developed and tested with v14.3.1
  \item \href{https://heasarc.gsfc.nasa.gov/fitsio/}{\texttt{cfitsio} library}, currently shipping with v4.3.0
  \item \href{https://github.com/scandum/quadsort}{\texttt{quadsort} library}, currently shipping with v1.2.1.2
  \item \href{https://github.com/DEShawResearch/random123}{\texttt{Random123} library}, currently using v1.14.0
  \item \href{https://github.com/MichaelJBerk/FontPopUp}{\texttt{FontPopUp} module}, currently using v1.0
\end{itemize}
HEALPix Viewer relies on \texttt{cfitsio} for basic file I/O, which I packaged as a Swift module (along with C HEALPix facilities) pre-configured for macOS, available \href{https://github.com/andrei-v-frolov/cfitsio}{on GitHub}. Source code for \texttt{quadsort} was patched for multi-threaded execution, and is included in \href{https://github.com/andrei-v-frolov/healpix-viewer}{HEALPix Viewer}. \texttt{Random123} by \cite{random123} is a header-only library, which does not require additional configuration. \texttt{FontPopUp} is a native Swift module, and is linked in XCode build configuration. All of the dependencies are included in the binary build, so no runtime dependencies exist apart from macOS SDK.

\section{Compliance and testing}
\label{sec:testing}

While HEALPix binary table read code was ported verbatim from original Fortran implementation and is reasonably safe, the entire metadata parsing code is written from scratch, and requires further testing and validation. All possible effort was made to comply with original \href{https://healpix.sourceforge.io/data/examples/healpix_fits_specs.pdf}{HEALPix specification}, but de-facto practices of major experiments do not necessarily follow it, and some allowances had to be made in interpretation. HEALPix Viewer is tested against WMAP DR5 data \citep{2013ApJS..208...20B}, Planck Legacy component-separated CMB and foreground data \citep{2020A&A...641A...1P, 2020A&A...641A...4P}, as well as frequency band maps \citep{2020A&A...641A...2P, 2020A&A...641A...3P}. It is also tested against maps output by \texttt{healpy}, for example, synthetic digital Earth elevation data at $15''$ resolution derived from (\href{http://www.viewfinderpanoramas.org/dem3.html}{de Ferranti et.\ al., 2020}), as illustrated in Fig.~\ref{fig:earth}, and output of my own code using Fortran HEALPix routines \citep{2020A&A...641A...7P}.

A major compliance issue is the current lack of partial sky data format support, which is documented in \href{https://healpix.sourceforge.io/data/examples/healpix_fits_specs.pdf}{HEALPix specification}, but is not widely used. Reputedly, Boomerang \citep{2000Natur.404..955D, 2003ApJS..148..527C} used this format, but I was unable to track down any actual maps for testing, so it remains unimplemented. If anybody has maps in this format, please get in touch with me!

Some unit tests were implemented for backbone functionality, in particular HEALPix routines imported from C, and color space conversions. GPU kernels are not as straightforward to test (requiring substantial effort to set up and import the results), so they currently remain in manual testing domain.

\section{Performance and benchmarks}
\label{sec:performance}

\begin{figure}
  \begin{center}
    \epsfig{file=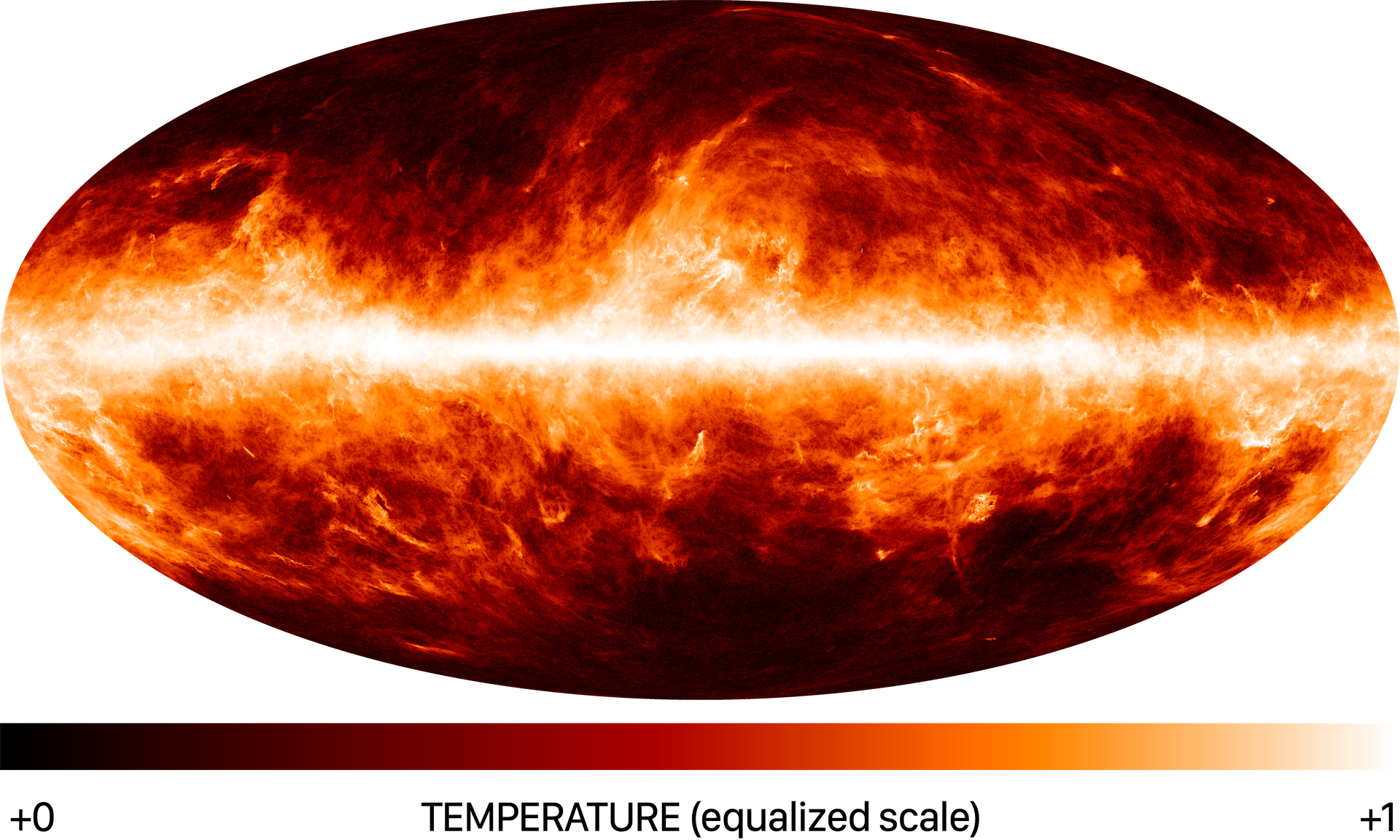, width=\columnwidth}
  \end{center}
  \caption{Planck 353GHz dust map (353GHz data with CMB and point source contributions subtracted) as rendered by HEALPix Viewer for export with 4x oversampling. Colorbar corresponds to equalized map, and is clamped to $[0,1]$.}
  \label{fig:353}
\end{figure}

\begin{table*}
  \begin{center}
    \begin{tabular}{c|rr|rr|rr|rr}
      & \textbf{Python} & \textbf{Viewer} & \textbf{Python} & \textbf{Viewer} & \textbf{Python} & \textbf{Viewer} & \textbf{Python} & \textbf{Viewer}\\
      & \multicolumn{2}{c|}{WMAP W-band} & \multicolumn{2}{c|}{Planck 857Ghz} & \multicolumn{2}{c|}{Earth DEM} & \multicolumn{2}{c}{Earth DEM}\\
      & \multicolumn{2}{c|}{$n_{\text{side}} = 512$} & \multicolumn{2}{c|}{$n_{\text{side}} = 2048$} & \multicolumn{2}{c|}{$n_{\text{side}} = 2048$} & \multicolumn{2}{c}{$n_{\text{side}} = 8192$}\\
      \hline
      Load map & 48.3ms & 48.2ms & 0.813s & 0.272s & 0.794s & 0.314s & 14.3s & 4.80s \\
      Rank map & 0.378s & 0.236s & 7.74s & 5.63s & 2.82s & 1.46s & 55.9s & 26.0s \\
      Transform & 8.7ms & 0.5ms & 0.165s & 10.7ms & 0.174s & 10.7ms & 2.92s & 0.157s \\
      Render PNG & 1.90s & 8.2ms & 2.00s & 8.1ms & 1.71s & 7.7ms & 1.72s & 8.3ms \\
      \hline
    \end{tabular}
  \end{center}
  \caption{Performance benchmarks for Python software stack and HEALPix Viewer on Apple M1 Max laptop with 64Gb RAM for four different datasets at three resolution levels. Times reported are wall time (for GPU from encode to command completion, which flushes the processing pipeline). }
  \label{tab:benchmark}
\end{table*}

From inception, the idea was to make an \textit{interactive} data visualization and exploration tool, which means data processing has to be fast enough to keep up with user inputs. That design goal was achieved in HEALPix Viewer. Its processing pipeline is vastly more efficient than tweaking \texttt{map2gif} parameters and re-rendering (which involves map load from disk every time \texttt{map2gif} is run), and outperforms CPU-bound solutions (such as \texttt{healpy} ran in an interactive envirnonment) by a large margin. In fact, all of the data processing except initial map load and ranking appears near-instantaneous to user, with map view responding to parameter changes in real time. Geometry changes involve only rendering at screen resolution (which is not that high compared to supported map size), and run at full 60fps at any map size tested. In fact, geometry rendering is so efficient it can be ran continuously without substantial power draw by the GPU. Triple buffering is employed to keep GPU render pipeline from starvation.

Despite being fast, the rendering pipeline is accurate to single precision (limited by 32-bit floats used in processing). HEALPix pixel boundaries and sphere geometry projections are accurate even at the highest zoom level (a factor of $1024$), unlike some implementation which use OpenGL vertex shaders (which usually involves approximating sphere geometry by polygons). Data statistics and moments are derived from down-sampled CDF (currently set to 4096 samples, but easily changed), and so are less precise than what could be achieved in a more sophisticated analysis. Exported image is in full 32-bit color supporting transparency, and oversampling up to a factor of $4$ is available to improve rendering quality of text fonts and noisy maps if desired. Typical output is shown in Figure~\ref{fig:353}.

The main user-facing delay is initial load of data from the HEALPix file, which is blocking and comes with indefinite progress indicator in the user interface. This is mostly limited by \texttt{cfitsio} implementation and disk access performance, and can range around $270-775$ms per \texttt{NESTED} single precision map at $\text{nside}=2048$ resolution (read from solid state disk on M1 Max laptop). Down-converting, reordering, and sign flips of polarization data are done in a single pass by a number of hard-coded C kernels for performance (Swift array implementation is fairly fast, but enforces bounds check on every access, which is quite noticeable overhead at data sizes involved). Initial load time is affected by data configuration, in particular \texttt{RING} to \texttt{NESTED} reordering can add up to $600$ms overhead per map. Further optimization is not feasible, as it would require rewriting large parts of \texttt{cfitsio}, which is a fairly monumental task. While sorting and ranking loaded maps can be resource-intensive, these operations are performed concurrently in the background, mitigating the processing delay experienced by the user (features which depend on ranked data get released as soon as data is available).

HEALPix Viewer (with \texttt{-O3} optimization of C code) was benchmarked against natively compiled Python v3.10 software stack with \texttt{healpy} v1.16.1 (as installed from \href{https://www.macports.org}{MacPorts} with default optimization level) for four different maps (WMAP W-band, Planck 2018 857GHz, and synthetic Earth elevation data at two resolutions as produced by \texttt{healpy}), with results reported in Table~\ref{tab:benchmark}. Both for Python and HEALPix Viewer, rendering times scale with output image size. The times reported here are for 3840 horizontal pixels bitmap rendered in Mollweide projection and encoded to PNG format. While original Fortran HEALPix implementation is quite impressive in performance, it is currently hampered by the lack of native Fortran compilers on Apple Arm architecture. Running it in emulation from Intel binaries would not do it justice, so it was excluded from comparison for now.

HEALPix Viewer code outperforms Python software stack across the board by a large margin (3-180x), both in CPU- and GPU-bound operations (which is surprising given amount of effort that went into optimizing Python libraries). Even ranking, which should be one of the most heavily optimized parts in most libraries, is considerably faster in HEALPix Viewer. I certainly cannot claim the credit for load and ranking performance (both of which are CPU-bound and use external libraries), but I can attest to \texttt{quadsoft}  being quite a bit faster than the default \texttt{scipy.stats.rankdata} algorithm, and perhaps raise a question of efficiency of \texttt{astropy.io.fits} input routines. GPU rendering of output image as implemented in HEALPix Viewer is vastly faster than \texttt{healpy/matplotlib} combo, but the HEALPix Viewer rendering code could be back-ported to Python using OpenCL or Metal backends to fix that.

\section{Current limitations}
\label{sec:limitations}

HEALPix Viewer is capable of dealing with extremely large maps without complexity associated with hierarchical data representation. Main limitation on the map size comes not from rendering pipeline performance, but physical size of VRAM and hard-coded limits on texture sizes in GPU libraries. As all the data gets loaded into GPU memory, it could easily exceed VRAM amount of consumer video cards. A single HEALPix map has $n_{\text{px}} = 12\, \text{nside}^2$ pixels, at $4$ bytes per pixel in a single precision buffer (of which $3$ are used). A color-mapped texture has $4n_{\text{px}}$ floating point RGBA values, at $4$ bytes per value for single precision. Total maximal amount of VRAM needed per map is then $336\, \text{nside}^2$ bytes, or $1.3$Gb per $\text{nside}=2048$ map, plus whatever memory needed for UI and output rendering (normally much less). Shared memory architectures like M1 and M2 handle this well, provided there's sufficient system memory. However, it may pose limitations on older hardware with separate GPU video cards (VRAM amounts of $4$Gb are common). Memory footprint can be significantly reduced by using 8-bit or 10-bit packed integer textures (at expense of slightly worse color gradation), which is available as an option, with required memory of $0.75$Gb per $\text{nside}=2048$ map.

Second limit comes from hard-coded maximal texture size for Metal libraries, which is $16384$ pixels, and limits nside of the map to $16384$ and output bitmap resolution to $16384/n_{\text{os}}$. This means that Simons Observatory maps are supported at full resolution ($\text{nside}=8192$), and maximal printed output size would be $2.8 \times 1.4$m at $150$dpi resolution. $\text{nside} = 8192$ maps were tested on commodity laptop used for development (M1 Max with $64$Gb system memory), and geometry rendering still runs at full 60fps. Data transforms at this resolution have noticeable lag, but are still quite responsive. Perceived performance could be improved by operating on downsampled proxy map during active user input phase (e.g.\ slider adjustment), with full resolution processed as soon as parameter values stop changing. The code was tested on both higher- and lower-end hardware as well, but permutations tested are limited to my hardware collection. Community feedback would be very welcome on this.

Multiple GPUs are currently not supported, but preference for a specific GPU can be chosen in Settings. While the data processing pipeline is trivially parallelizable, bookkeeping and memory management details required for GPU load balancing would have to be hand-managed and would slow down the development considerably. With current hardware presenting itself as a single GPU, this is not a priority, but could present an issue on older machines with multiple discrete or external GPUs. By default, the first GPU on the system list is used for rendering, with specific class preference specifiable in Settings.

HEALPix viewer binary is built against macOS 12 libraries, with exception of data statistics overlay which requires plotting libraries which only became available in macOS 13 (and is disabled on earlier versions). While most of the GPU kernels can be easily ported to OpenCL to be ran on other operating systems, the rest of the source code is tightly coupled to SwiftUI libraries, and porting user interface code would involve complete re-write. For this reason, support for Windows and Linux is not planned.

\section{Development roadmap}
\label{sec:development}

Current development is focusing on quality of life improvements (map render state persistence, copy and paste of various settings, ability to save/load entire workspace as a JSON file, etc.), implementing GUI for desirable options (proper Settings dialog, custom gradient editor, etc.), as well as developing novel visualization tools (false color maps, on-the-fly component separation, visualization of directional fields via LIC, etc.). Implementing an interactive calculator for HEALPix data is currently under investigation. Once the feature set is frozen for v2.0 release, I plan to add an interactive walk-through of the features (available at first launch). Proper help bundle will be eventually added to the distribution.

Among the more interesting forthcoming tools is on-the-fly component separation. Current component separation strategies rely either on fitting emission frequency spectrum model, e.g.\ Gibbs samplers like Commander \citep{2004PhRvD..70h3511W, 2008ApJ...676...10E}, or constructing minimal variance estimators, e.g.\ WMAP ILC \citep{2003ApJS..148....1B}, Planck NILC \citep{2011MNRAS.418..467R}, SEVEM \citep{2003MNRAS.345.1101M}, and SMICA \citep{2003MNRAS.346.1089D, 2008arXiv0803.1814C}. Minimal variance estimators generally require data reduction, which is somewhat involved on GPU, while Gibbs samplers can operate per pixel, and are perfect for parallelization. Preliminary implementation of simplified Commander-like component separator with no beam or point source support and fixed emission spectra parameters benchmarks at about 30ms for three $\text{nside}=2048$ map inputs. This level of performance would allow interactive estimation of various emission components, and visually facilitate understanding of how emission parametrization affects the component separation. More complicated spatial priors could be introduced, with scale-invariant one being the most interesting physically while still allowing efficient implementation on GPU via multigrid solver \citep{Brandt:1977, 2020A&A...641A...7P}.

As a long-term goal, integration with existing online astronomy databases is being considered. This could include support for HiPS \citep{2015A&A...578A.114F}, direct integration with datasets available through LAMBDA \citep{2019arXiv190508667A} or query-based databases like \href{https://www.cadc-ccda.hia-iha.nrc-cnrc.gc.ca/en/}{CADC}. At the moment, main limitation is the available development time, so these features don't have a specific target time frame for implementation.

\section{Conclusions}
\label{sec:conclusions}

Cosmic microwave background is one of the pillars of precision cosmology, and a number of new high-resolution surveys are being built or planned. Multi-band temperature and polarization maps are needed to remove astrophysical backgrounds and reconstruct primordial CMB fluctuations. CMB experiments not only constrain a handful of cosmological parameters, but serve as a window to early universe physics (in particular, the search for primordial gravitational waves is currently under way with attempts to detect B-mode polarization) and provide astrophysical insights with reconstructed foreground maps (for example, information on magnetic field of our own galaxy). Full or partial sky maps are produced from time-ordered data by sophisticated mapping algorithms, and main data analysis is carried out on this reduced data set. Storing or processing it requires a way to pixelize the sphere, and HEALPix (Hierarchical Equal Area isoLatitude Pixelation) is widely used for that purpose.

This paper presents a new interactive visualization application developed for macOS using latest software and technology, which I called HEALPix Viewer for the lack of a better name. It is fully GPU accelerated, and can handle extremely large maps (which is a common trend in future experiments, for example Simons Observatory would produce almost giga-pixel maps). Software design was done from scratch and uses modern language features of Swift to full advantage, making the code modular and easy to maintain. Data pipeline was designed with performance in mind, and is fast enough to process and render $\text{nside}=8192$ maps in real time on a laptop, given enough memory. Overall performance is an order or two of magnitude improvement over existing Python software stack, which enables disruptive paradigm change in data analysis and visualization. User interface makes many common tasks easy and convenient, for example rendered maps can be dropped directly into a Keynote or Powerpoint presentation. HEALPix Viewer is distributed as an universal application that can natively run on Intel and Apple Silicon hardware for \href{https://apps.apple.com/app/healpix-viewer/id1660836459}{easy installation} if you just want to use it, and has complete \href{https://github.com/andrei-v-frolov/healpix-viewer}{source code} available if you want to add new features or use the code in other projects.

HEALPix Viewer represents my take on what fast imaging pipeline for HEALPix data could be, and is designed to be easily extensible. Development is ongoing, and further improvements could be made in the future. In particular, visualization of polarization and vector fields via line integral convolution \citep{convolution} is planned. Feedback and feature requests are welcome, and should be directed to the author.

\section*{Acknowledgements}
This work was supported in part by NSERC Discovery Grant \textsl{``Testing fundamental physics with B-modes of Cosmic Microwave Background anisotropy''}. The author thanks Dick Bond, Sigurd K. N{\ae}ss, and Douglas Scott for stimulating discussions, and everyone who participated in beta testing.

\appendix

\section{Projections of a sphere}
\label{sec:projection}

\subsection{Mollweide projection}
Mollweide projection (commonly used in CMB literature) maps an entire sphere onto an ellipse of $2$:$1$ aspect ratio. It is an equal-area pseudocylindrical projection which keeps iso-latitude lines straight and horizontal, at expense of shape distortion near the boundaries. Inverse transformation from Cartesian plane $(x,y)$ to spherical coordinates $(\theta,\phi)$ is
\begin{equation}
  \xi = \arcsin y,\hspace{0.3em}
  \theta = \arccos\frac{2\xi + \sin(2\xi)}{\pi},\hspace{0.3em}
  \phi = \frac{\pi x}{2\cos\xi}.
\end{equation}
Projection bounds are set by $|y| \le 1$ and $|\phi| \le \pi$, with projection extent $|x| \le 2$ and $|y| \le 1$.

\subsection{Hammer projection}
Visually similar to Mollweide projection, Hammer projection trades reduced shape distortion for iso-latitude lines being curved. It is also equal area and maps an entire sphere onto an ellipse of $2$:$1$ aspect ratio. Inverse transformation from Cartesian plane $(x,y)$ to spherical coordinates $(\theta,\phi)$ is
\begin{equation}  \textstyle
    q = 1 - \frac{x^2}{16} - \frac{y^2}{4},\hspace{0.0em}
    \theta = \arccos\left(q^{\frac{1}{2}}y\right),\hspace{0.0em}
    \phi = 2 \arctan\frac{q^{\frac{1}{2}}x}{2(2q-1)}.
\end{equation}
Projection bounds are set by $q \ge 1/2$, with projection extent $|x| \le \sqrt{8}$ and $|y| \le \sqrt{2}$.

\subsection{Lambert projection}
Also known as Lambert azimuthal projection, it maps an entire sphere onto a disk. It is also an equal-area projection, with inverse transformation from Cartesian plane $(x,y)$ to unit vector $\vec{n}$
\begin{equation}
    q = 1 - (x^2 + y^2)/4,\hspace{1em}
    \hat{\vec{n}} = \left[2q-1, q^{\frac{1}{2}} x, q^{\frac{1}{2}} y\right].
\end{equation}
Projection bounds are set by $q \ge 0$, with projection extent $|x| \le 2$ and $|y| \le 2$.

\subsection{Orthographic (isometric) projection}
Isometric, or perhaps more appropriately orthographic, projection is the view of a sphere by an observer at infinity. It maps half of a sphere onto a disk. This is the most familiar projection, as it corresponds to looking at a physical sphere. Inverse transformation from Cartesian plane $(x,y)$ to unit vector $\vec{n}$ is
\begin{equation}
    q = 1 - x^2 - y^2,\hspace{1em}
    \hat{\vec{n}} = \left[q^{\frac{1}{2}}, x, y\right].
\end{equation}
Projection bounds are set by $q \ge 0$, with projection extent $|x| \le 1$ and $|y| \le 1$.

\subsection{Stereographic projection}
Stereographic projection is defined by mapping the sphere to a tangent plane in the direction of a vector extended from the opposite point. It is conformal, projects circles on a sphere to circles on a plane, and maps an entire sphere onto an infinite plane. Inverse transformation from Cartesian plane $(x,y)$ to unit vector $\vec{n}$ is
\begin{equation}
    q = 1 + (x^2 + y^2)/4,\hspace{1em}
    \hat{\vec{n}} = \left[2,x,y\right]/q - \left[1,0,0\right].
\end{equation}

\subsection{Gnomonic projection}
Gnomonic projection is defined by mapping the sphere to a tangent plane in the direction of radial vector. It maps great circles to straight lines (meaning projected straight lines are geodesic), and maps half of a sphere onto an infinite plane. Inverse transformation from Cartesian plane $(x,y)$ to unit vector $\vec{n}$ is
\begin{equation}
    \hat{\vec{n}} = \left[1,x,y\right]/\sqrt{1+x^2+y^2}.
\end{equation}

\subsection{Mercator projection}
Mercator projection is a cylindrical projection quite often used in cartography and navigation. It is conformal and maps the entire sphere onto an infinite strip. Inverse transformation from Cartesian plane $(x,y)$ to spherical coordinates $(\theta,\phi)$ is
\begin{equation}
    \phi = x,\hspace{1em}
    \theta = \frac{\pi}{2} - \arctan(\sinh y).
\end{equation}
Projection bounds are set by $|\phi| \le \pi$, with infinite extent in $y$ direction.

\subsection{Cartesian (cylindrical) projection}
Cartesian, or plate carr\'ee, projection is a trivial mapping of spherical coordinates onto a plane. It is neither equal-area nor conformal, but it maps parallels and meridians into straight lines. It maps an entire sphere onto a rectangle of $2$:$1$ aspect ratio. Distortion near poles is extreme. Inverse transformation from Cartesian plane $(x,y)$ to spherical coordinates $(\theta,\phi)$ is
\begin{eqnarray}
    \phi = x,\hspace{1em}
    \theta = \frac{\pi}{2} - y.
\end{eqnarray}
Projection bounds are set by $|\phi| \le \pi$, $0\le\theta\le\pi$, with corresponding extent $|x| \le \pi$ and $|y| \le \pi/2$.

\subsection{Werner projection}
Werner projection is an equal-area pseudoconic projection which maps the entire sphere into a heart shape. Least commonly used of the projections listed here, it nevertheless was known from circa 16th century. Inverse transformation from Cartesian plane $(x,y)$ to spherical coordinates $(\theta,\phi)$ is
\begin{eqnarray}
    \rho = \sqrt{x^2+y^2},\hspace{1em}
    \phi = \frac{\rho}{\sin \rho}\,\atan(x,-y),\hspace{1em}
    \theta = \rho.
\end{eqnarray}
Projection bounds are set by $|\phi| \le \pi$, $0\le\theta\le\pi$, with projection extent being asymmetrical in $y$ direction and transcendental. After shifting $y$ coordinate by $1.111983413$ to center the image, the extent is almost exactly at $1$:$1$ aspect ratio, with $|x|\le 2.021610497$ and $|y| \le 2.029609241$. It is implemented because it looks cool.

\section{Representations of rotation group}
\label{sec:representation}

\subsection{Euler angles}
\label{sec:rotation}

Rotation of a sphere which brings a viewpoint with a given latitude $\vartheta$, longitude $\phi$, and azimuth $\psi$ to the direct line of sight can be parametrized as a composition of rotations in $xy$, $xz$, and $yz$ planes. These are usually known as Euler angles, with
\begin{eqnarray}
  \mathbb{R}_{xz} &=& \left[\begin{array}{rrr}
      \cos\vartheta & 0 & -\sin\vartheta\\
      0 & 1 & 0\\
      \sin\vartheta & 0 & \cos\vartheta\\
    \end{array}\right],\\
  \mathbb{R}_{xy} &=& \left[\begin{array}{rrr}
      \cos\phi & -\sin\phi & 0\\
      \sin\phi & \cos\phi & 0\\
      0 & 0 & 1\\
    \end{array}\right],\\
  \mathbb{R}_{yz} &=& \left[\begin{array}{rrr}
      1 & 0 & 0\\
      0 & \cos\psi & -\sin\psi\\
      0 & \sin\psi & \cos\psi\\
    \end{array}\right].
\end{eqnarray}
Note that some of the signs are different from the usual notation, because we are interested in rotation which brings a given location into a fixed position, not the other way around. Order of rotations matters, as they do not commute. The composite rotation matrix is
\begin{equation}\label{eq:rot:euler}
  \mathbb{R} = \mathbb{R}_{xy}\, \mathbb{R}_{xz}\, \mathbb{R}_{yz}
    = \left[\begin{array}{rrr}
          R_{11} & R_{12} & R_{13}\\
          R_{21} & R_{22} & R_{23}\\
          R_{31} & R_{32} & R_{33}
        \end{array}\right],
\end{equation}
with matrix elements being
\begin{eqnarray}
   R_{11} &=& \cos\vartheta \cos\phi,\label{eq:rot:euler:11}\\
   R_{12} &=& -\sin\vartheta \cos\phi \sin\psi - \sin\phi \cos\psi,\label{eq:rot:euler:12}\\
   R_{13} &=& -\sin\vartheta \cos\phi \cos\psi + \sin\phi \sin\psi,\label{eq:rot:euler:13}\\
   R_{21} &=& \cos\vartheta \sin\phi,\label{eq:rot:euler:21}\\
   R_{22} &=& -\sin\vartheta \sin\phi \sin\psi + \cos\phi \cos\psi,\label{eq:rot:euler:22}\\
   R_{23} &=& -\sin\vartheta \sin\phi \cos\psi - \cos\phi \sin\psi,\label{eq:rot:euler:23}\\
   R_{31} &=& \sin\vartheta,\label{eq:rot:euler:31}\\
   R_{32} &=& \cos\vartheta \sin\psi,\label{eq:rot:euler:32}\\
   R_{33} &=& \cos\vartheta \cos\psi.\label{eq:rot:euler:33}
\end{eqnarray}
It should be remembered that in Swift, vectorized matrix types like \code{float3x3} are stored column-wise, with indexing starting from zero, so matrix element $R_{31}$ would actually be referred to as \code{R[0,2]} in the code, and so on. Actual rotation matrix is evaluated as a product \ref{eq:rot:euler} in the code, not long-form expressions of (\ref{eq:rot:euler:11}--\ref{eq:rot:euler:33}), which is vectorized and more efficient.

Euler angles can be extracted from a given rotation matrix $\mathbb{R}$
\begin{eqnarray}
  \vartheta &=& \arcsin R_{31},\label{eq:angle:theta}\\
  \phi &=& \atan(R_{21},R_{11}),\label{eq:angle:phi}\\
  \psi &=& \atan(R_{32},R_{33}),\label{eq:angle:psi}
\end{eqnarray}
although one has to be mindful of the branches. The $\arcsin$ expression (\ref{eq:angle:theta}) is meant to have $|\vartheta|< \pi/2$, corresponding to spherical coordinate $\theta = \pi/2-\vartheta$ range of $0<\theta<\pi$. The last two expressions go degenerate when $|\vartheta|=\pm\pi/2$ (i.e.\ at the poles), for which case the rotation matrix reads
\begin{eqnarray}    
  \mathbb{R} &=& \left[\begin{array}{rrr}
          0 &
          - \sin(\phi\pm\psi) &
          \mp \cos(\phi\pm\psi)\\
          0 &
          \cos(\phi\pm\psi) &
          \mp \sin(\phi\pm\psi)\\
          \pm 1 & 0 & 0
        \end{array}\right]. \nonumber
\end{eqnarray}
Longitude and azimuth cannot be distinguished at the poles, and only their combination
\begin{equation}
  \phi\pm\psi = \atan\left(-R_{12},R_{22}\right)
\end{equation}
is determined. For definiteness, one can choose $\psi=0$ there.

\subsection{Generators of rotation}
\label{sec:generator}

Special orthogonal transformations form a Lie group, and as such are representable by corresponding Lie algebra generators. A rotation matrix $\mathbb{R}$ is thus representable by exponent of an antisymmetric matrix $\mathbb{W}$
\begin{equation}\label{eq:generator:exp}
  \mathbb{R} = \exp[\mathbb{W}] \equiv \sum\limits_{n=0}^{\infty} \frac{\mathbb{W}^n}{n!}.
\end{equation}
In three dimensions, 3x3 antisymmetric matrix $\mathbb{W}$ has only 3 independent components, and can be written as a dual of three-vector $\vec{w}$
\begin{equation}
  \mathbb{W} \equiv *\vec{w} = \left[\begin{array}{ccc}
    0 & -w_z & w_y \\
    w_z & 0 & -w_x \\
    -w_y & w_x & 0 \\
  \end{array}\right].
\end{equation}
A specific property of 3x3 antisymmetric matrices which simplifies things a lot is that Krylov space has low dimension, since 
\begin{equation}
  \mathbb{W}^3 = -||\mathbb{W}||^2\, \mathbb{W},
\end{equation}
where matrix norm is defined as
\begin{equation}
  ||\mathbb{W}||^2 \equiv -\frac{1}{2}\, \textrm{Tr}\, \mathbb{W}^2 = |\vec{w}|^2.
\end{equation}
Resumming power series in (\ref{eq:generator:exp}), one obtains
\begin{equation}
  \mathbb{R} = \exp[\mathbb{W}] = \mathbb{I} + \frac{\sin w}{w}\, \mathbb{W} + 2\,\frac{\sin^2 \frac{w}{2}}{w^2}\, \mathbb{W}^2,
\end{equation}
where $\mathbb{I}$ is identity matrix and $w = |\vec{w}|$. Conversely, one can read off generator vector from a given rotation matrix $\mathbb{R}$ by
\begin{equation}
  \vec{w} = \atan\left(s,{\textstyle\frac{1}{2}} (t-1)\right)\, \hat{\vec{s}},
\end{equation}
where
\begin{eqnarray}
  \vec{s} &=& *\mathbb{R} \equiv {\textstyle\frac{1}{2}} \left[R_{32}-R_{23}, R_{13}-R_{31}, R_{21}-R_{12}\right],\\
  t &=& \textrm{Tr}\, \mathbb{R} = R_{11} + R_{22} + R_{33}.
\end{eqnarray}
Generator of rotation $\vec{w}$ represents rotation by an angle $w$ around the axis $\hat{\vec{w}}$, and is defined up to an integer multiple of $2\pi$ in length. When setting the target rotation generator, one should pick the value closest to the current one among possible equivalent choices to ensure the shortest rotation between the two.

\section{Sphere dynamics}
\label{sec:integrator}

Transition between different viewpoints specified by user is animated by solving a damped simple harmonic oscillator equation for rotation generator $\vec{w}$, asymptoting to target location $\vec{t}$
\begin{equation}
  \ddot{\vec{w}} + \gamma \dot{\vec{w}} + \kappa(\vec{w} - \vec{t}) = 0.
\end{equation}
This can be readily evolved by second-order accurate operator splitting scheme
\begin{eqnarray}
  \vec{w} &\mapsto& \vec{w} + \dot{\vec{w}} \, {\textstyle\frac{dt}{2}}\label{eq:integrator:kick:1}\\
  \dot{\vec{w}} &\mapsto& \dot{\vec{w}} - \Big[\gamma \dot{\vec{w}} + \kappa(\vec{w} - \vec{t}) \Big]\, dt\\
  \vec{w} &\mapsto& \vec{w} + \dot{\vec{w}} \, {\textstyle\frac{dt}{2}}\label{eq:integrator:kick:2}
\end{eqnarray}
commonly used and known by many names (e.g.\ leapfrog and kick-drift). It happens to be symplectic for Hamiltonian equations of motion (i.e.\ a canonical transformation, albeit for approximate Hamiltonian), and the order can be increased by forming operator sandwiches with particularly selected time steps, so higher order commutators would vanish. Among those, 6-th order symplectic scheme first derived by \cite{1990PhLA..150..262Y} is still cheap and very convenient to implement
\begin{equation}
  H_6(\Delta t) = \prod\limits_{-3\le i\le 3\phantom{-}} H_2\left(\alpha_{|i|} \Delta t\right),
\end{equation}
where $H_2$ is the second order evolution step (\ref{eq:integrator:kick:1}--\ref{eq:integrator:kick:2}), and coefficients $\alpha_{|i|}$ are specifically chosen to be (in quad precision)
\begin{eqnarray}
  \alpha_0 &=& \phantom{-}1.315186320683911218884249728239,\\
  \alpha_1 &=&           -1.177679984178871006946415680964,\\
  \alpha_2 &=& \phantom{-}0.235573213359358133684793182979,\\
  \alpha_3 &=& \phantom{-}0.784513610477557263819497633866.
\end{eqnarray}

\section{Inverse error function}
\label{sec:erfinv}

To transform uniform distribution into a Gaussian one, one needs to compute an inverse error function. Since this is not a part of standard GPU libraries, a workable implementation is needed. An approximation derived in \cite{erfinv} is as follows
\begin{equation}
  \text{erf}^{-1}(x) = \left\{
    \begin{array}{cc}
      x p_1(w), & w < 5\\
      x p_2(w), & w \ge 5
    \end{array}
  \right.,
\end{equation}
where $p_1$ and $p_2$ are polynomials of order 8, and
\begin{equation}
  w = -\ln(1-x^2).
\end{equation}
Polynomial $p_1$ (in Horner form) is evaluated as
\begin{verbatim}
  w = w - 2.5;
  p =  2.81022636e-08;
  p =  3.43273939e-07 + p*w;
  p = -3.5233877e-06  + p*w;
  p = -4.39150654e-06 + p*w;
  p =  0.00021858087  + p*w;
  p = -0.00125372503  + p*w;
  p = -0.00417768164  + p*w;
  p =  0.246640727    + p*w;
  p =  1.50140941     + p*w;
\end{verbatim}
while polynomial $p_2$ (actually of argument $\sqrt{w}$) is
\begin{verbatim}
  w = sqrt(w) - 3.0;
  p = -0.000200214257;
  p =  0.000100950558 + p*w;
  p =  0.00134934322  + p*w;
  p = -0.00367342844  + p*w;
  p =  0.00573950773  + p*w;
  p = -0.0076224613   + p*w;
  p =  0.00943887047  + p*w;
  p =  1.00167406     + p*w;
  p =  2.83297682     + p*w;
\end{verbatim}
The approximation is good down to last digit of single precision floats for entire range of argument values, produces the right asymptotics, and is quite cheap to evaluate (involving only 8 multiply-accumulate operators).

\section{Perceptually uniform color spaces}
\label{sec:color}

\subsection{Color representation basics}
Computer images are commonly represented by a triplet of color values (red, green, and blue) and optional opacity value (alpha) per pixel, with multitude of encoding and compression schemes employed by different image formats. On most GPUs, RGBA packed pixel values are natively supported by hardware at different bit width and precision, including floating point types. For storage, images are typically quantized to 8 bits per component, except for special applications where increased tonal gradation or dynamic range are important (as in high-end photographic and video productions).

The choice of three basis vectors to represent visible colors is motivated by properties of human vision, with the eye having three different color receptors, plus additional low-illumination one mostly insensitive to color. Thus in image acquisition, the full spectral response is usually integrated with three particular band-pass filters (although four are occasionally used to improve perceived color reproduction, for example Fujifilm Reala film emulsion had fourth cyan-sensitive layer). Native output of CCD and CMOS sensors is linearly proportional to incident light intensity (i.e.\ photon count), however in computer imaging a power law with exponent of approximately $0.45$ is applied to sampled values for historical reasons (CRTs had non-linear intensity response to driving voltage), which also helps with quantization appearing perceptually uniform. The choice of the color basis vectors (known as primaries), as well as non-linear transfer curves defines a color space, and image RGB values are essentially just coordinates on that manifold.

\begin{figure}
  \begin{center}
    \epsfig{file=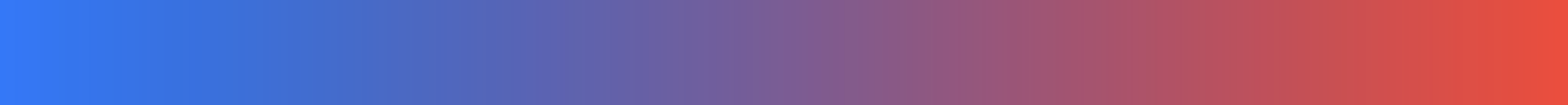, width=\columnwidth}\medskip\\
    \epsfig{file=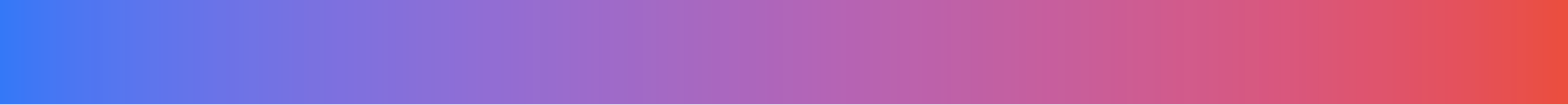, width=\columnwidth}\medskip\\
    \epsfig{file=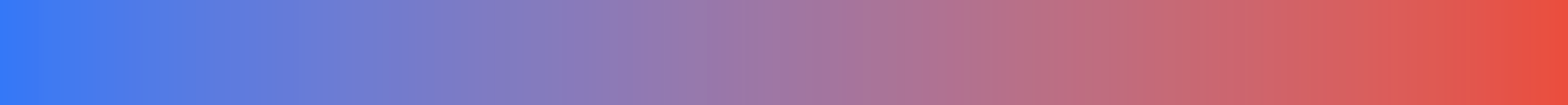, width=\columnwidth}
  \end{center}
  \caption{Linear interpolation between blue and red in device color space (sRGB), linear sRGB space, and perceptually uniform  color space (okLab). First transition has noticeable brightness change, the second one has unexpected strong hue change in purple direction, while the last one appears to be OK.}
  \label{fig:perception}
\end{figure}

\subsection{Perceptual color metrics}
As straightforward as integrating down with three band-pass filters to obtain a representation of color might sound, human vision is substantially more complex in nature, and \textit{does not} perceive linear intensity differences as uniform. In other words, all color spaces come outfitted with a metric reflecting perceptual differences. As an additional complication, human vision is very efficient at nulling out perceived differences in overall illumination spectra, leading to necessity of remapping absolute ``white'' values in color reproduction. A lot of effort has been put into construction of color spaces which would have Cartesian perceptual color difference metric, most notably \href{https://en.wikipedia.org/wiki/CIELAB_color_space}{CIELab}.

If perceptual color metric is ignored, mixing or linear interpolation between different colors might have unexpected results, which is most noticeable in red-blue transitions, as illustrated in Fig.~\ref{fig:perception}. When mixing colors, HEALPix Viewer implements straight linear combination, linear remapping of white point, and blending in perceptually uniform color space as user-specifiable strategies. These lead to noticably different results.

\subsection{OKLab color space}
\label{sec:oklab}

While many color spaces have been developed and proposed to address peculiarities of human vision (e.g.\ \href{https://en.wikipedia.org/wiki/CIELAB_color_space}{CIELab} and substantially more complex \href{https://en.wikipedia.org/wiki/CIECAM02}{CIECAM02}), we opted for a relative newcomer (\href{https://bottosson.github.io/posts/oklab/}{okLab}) as a colorspace of choice. It is very efficient to implement on GPU (consisting of a matrix transform, power law, followed by a second matrix transform), and is as good (if not better) in perceived color accuracy as much more complicated constructs.

Transformation from linear sRGB space to okLab is
\begin{equation}
  \left[\begin{array}{c} l\\ m\\ s\\\end{array}\right] = \mathbb{M}_1  \left[\begin{array}{c} R\\ G\\ B\\\end{array}\right],\hspace{1em}
  \left[\begin{array}{c} L\\ a\\ b\\\end{array}\right] = \mathbb{M}_2  \left[\begin{array}{c} l^{\frac{1}{3}}\\ m^{\frac{1}{3}}\\ s^{\frac{1}{3}}\end{array}\right],
\end{equation}
where
\begin{equation}
  \mathbb{M}_1 = \left[\scriptsize
  \begin{array}{ccc}
    +0.4122214708 & +0.5363325363 & +0.0514459929\\
    +0.2119034982 & +0.6806995451 & +0.1073969566\\
    +0.0883024619 & +0.2817188376 & +0.6299787005
  \end{array}
  \right],
\end{equation}
\begin{equation}
  \mathbb{M}_2 = \left[\scriptsize
  \begin{array}{ccc}
    +0.2104542553 & +0.7936177850 & -0.0040720468\\
    +1.9779984951 & -2.4285922050 & +0.4505937099\\
    +0.0259040371 & +0.7827717662 & -0.8086757660
  \end{array}
  \right].
\end{equation}

\subsection{Gamut mapping}
\label{sec:gamut}

\begin{figure}
  \begin{center}
    \epsfig{file=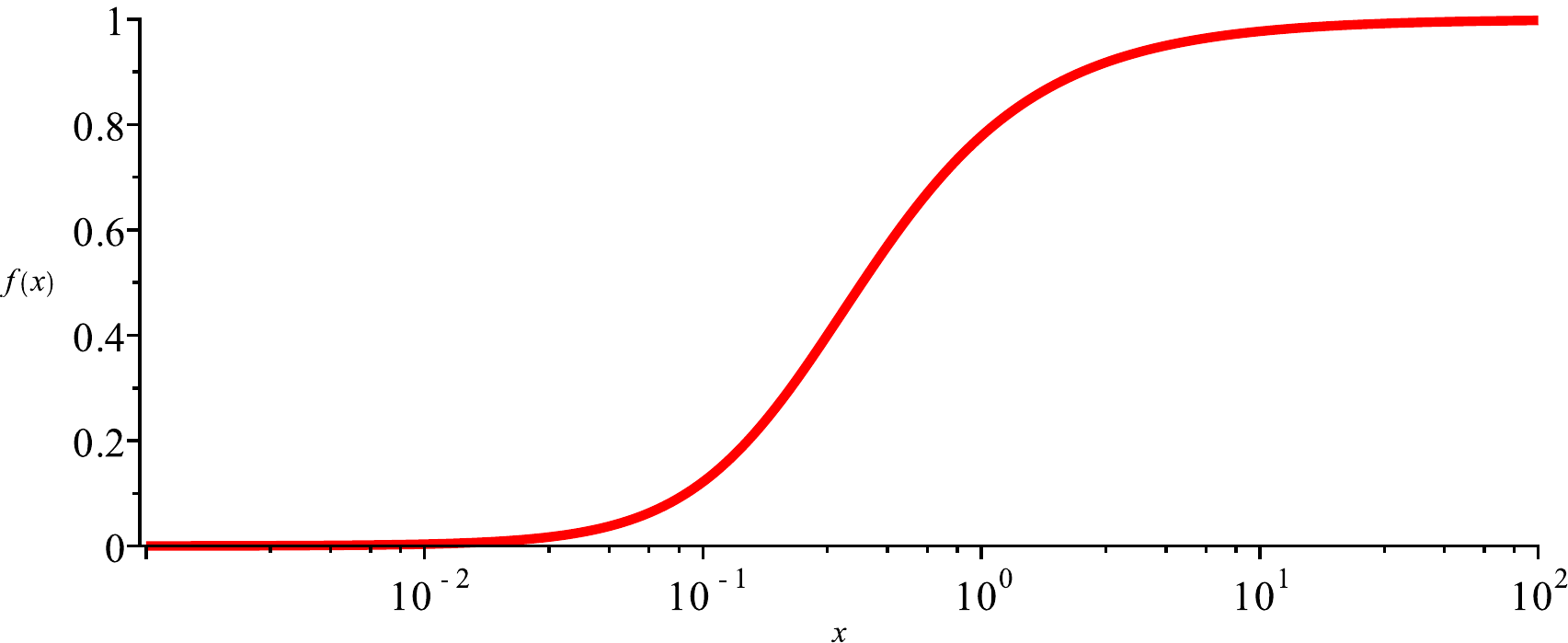, width=\columnwidth}
  \end{center}
  \caption{Filmic transfer curve used to compress the color component values into a unit range.}
  \label{fig:filmic}
\end{figure}

When generating false color maps, one could encounter colors not representable by the output device or storage format, e.g.\ negative RGB values, or values exceeding the allowed upper limit. Simple clipping to bring the values in range (as done for single-value color palette lookups) could lead to strange color shifts, so more sophisticated mapping of out-of-range values is often employed. HEALPix Viewer implements the Academy Color Encoding System (ACES) \href{https://docs.acescentral.com/guides/rgc-user/}{reference gamut compression} to deal with over-saturated colors, and approximate ACES \href{https://knarkowicz.wordpress.com/2016/01/06/aces-filmic-tone-mapping-curve/}{filmic curve}
\begin{equation}
  f(x) = \frac{2.43}{2.51}\frac{(2.51x+0.03)x}{(2.43x+0.59)x+0.14}
\end{equation}
to bring color component values into a unit range, as illustrated in Fig.~\ref{fig:filmic}. Other options (such as BBC/NHK \href{https://en.wikipedia.org/wiki/Hybrid_logÐgamma}{hybrid log-gamma} transfer curves) were explored, but the above choice seems to produce the best visual results.

\bibliographystyle{model2-names}
\bibliography{paper,healpix,planck,compsep,cmb,xray}

\end{document}